\documentclass[12pt]{article}
\usepackage{latexsym}
\usepackage{bbm}
\usepackage{amsopn}
\usepackage{amsfonts}
\usepackage{amssymb}
\usepackage{amsthm}
\usepackage{amsmath}

\renewcommand{\theequation}{\arabic{section}.\arabic{equation}}

\newtheorem{Theorem}{Theorem}[section]
\newtheorem{Proposition}[Theorem]{Proposition}
\newtheorem{Lemma}[Theorem]{Lemma}
\newtheorem{Corollary}[Theorem]{Corollary}
\newtheorem{Definition}[Theorem]{Definition}

\newcommand{\tree}{\mathcal{T}}

\begin{document}


\title{\bf Charge Superselection Sectors for
QCD on the Lattice}

\author{
    J. Kijowski \\
    Center for Theoretical  Physics, Polish Academy of Sciences\\
    al. Lotnik\'ow 32/46, 02-668 Warsaw, Poland\\
    \ \\
    G. Rudolph \\
    Institut f\"ur Theoretische Physik, Universit\"at Leipzig\\
    Augustusplatz 10/11, 04109 Leipzig, Germany\\
    }

\maketitle


\begin{abstract}
We study quantum chromodynamics (QCD) on a finite lattice
$\Lambda$ in the Hamiltonian approach. First, we present the field
algebra ${\mathfrak A}_{\Lambda}$ as comprising a gluonic part,
with basic building block being the crossed product $C^*$-algebra
$C(G) \otimes_{\alpha} G$, and a fermionic (CAR-algebra) part
generated by the quark fields. By classical arguments, ${\mathfrak
A}_{\Lambda}$ has a unique (up to unitary equivalence) irreducible
representation. Next, the algebra ${\mathfrak O}^i_{\Lambda}$ of
internal observables is defined as the algebra of gauge invariant
fields, satisfying the Gauss law. In order to take into account
correlations of field degrees of freedom inside $\Lambda$ with the
``rest of the world'', we have to extend ${\mathfrak
O}^i_{\Lambda}$ by tensorizing with the algebra of gauge invariant
operators at infinity. This way we construct the full observable
algebra ${\mathfrak O}_{\Lambda}\, .$ It is proved that its
irreducible representations are labelled by ${\mathbb Z}_3$-valued
boundary flux distributions. Then, it is shown that there exist
unitary operators (charge carrying fields), which intertwine
between irreducible sectors leading to a classification of
irreducible representations in terms of the ${\mathbb Z}_3$-valued
global boundary flux. By the global Gauss law, these 3
inequivalent charge superselection sectors can be labeled in terms
of the global colour charge (triality) carried by quark fields.
Finally, ${\mathfrak O}_{\Lambda}$ is discussed in terms of
generators and relations.
\end{abstract}

\newpage

\vspace{0.5cm}

\tableofcontents

\newpage


\setcounter{equation}{0}
\section{Introduction}


In a series of papers, we have started to analyze the
non-perturbative structure of gauge theories, with the final aim
being the formulation and investigation of gauge models purely in
terms of observables. To start with one should clarify basic
structures like that of the field algebra, the observable algebra
and the superselection structure of the Hilbert space of physical
states. It is well-known that the standard Doplicher-Haag-Roberts
theory \cite{DHR}, \cite{DR} for models, which do not contain
massless particles, does not apply here. Nonetheless, there are
interesting partial results within the framework of general
quantum field theory both for quantum electrodynamics (QED) and
for non-abelian models, see \cite{Bu}, \cite{JF}, \cite{SW1} and
\cite{S}.

To approach the problem in a rigorous way, we put the system on a
finite (regular cubic) lattice and formulate the model within the
Hamiltonian approach. For basic notions concerning lattice gauge
theories (including fermions) we refer to \cite{Seiler} and
references therein. Within the finite lattice context, we have
analyzed the structure of the observable algebra both for
spinorial and scalar QED and we have shown that the physical
Hilbert space decomposes into a direct sum of superselection
sectors labelled by the total electric charge, see \cite{KRT},
\cite{KRS} and \cite{KRS1}. Finally, of course, one wants to
construct the continuum limit. In full generality, this is an
extremely complicated problem of constructive field theory.
However, the results obtained until now suggest that there is some
hope, to control the thermodynamic limit, see \cite{KRT} for a
heuristic discussion. We also mention that for simple toy models,
these problems can be solved, see \cite{FM}.

In \cite{KR} we have started to investigate quantum chromodynamics
(QCD) within the above framework. In particular, we have analyzed
the global Gauss law and the notion of global colour charge
(triality). Comparing with QED, the notion of global charge in QCD
is much more complicated, according to the fact that the local
Gauss law is neither built from gauge invariant operators nor is
it linear. We have shown that one can extract from the local Gauss
equation of QCD a gauge invariant, additive law for operators with
eigenvalues in ${\mathbb Z}_3$, (the center of $SU(3)$). This
implies -- as in QED -- a gauge invariant conservation law: The
global ${\mathbb Z}_3$-valued colour charge is equal to a
${\mathbb Z}_3$-valued gauge invariant quantity obtained from the
color electric flux at infinity.

We stress that within lattice gauge theories, the notion of colour
charge is already implicitly contained in a paper by Kogut and
Susskind, see \cite{Kogut}. In particular, Mack \cite{Mack} used
it to propose a certain (heuristic) scheme of colour screening and
quark confinement, based upon a dynamical Higgs mechanism with
Higgs fields built from gluons. For similar ideas we also refer to
papers by `t Hooft, see \cite{tHooft} and references therein. This
concept was also used in a paper by Borgs and Seiler \cite{Borgs},
where the confinement problem for Yang-Mills theories with static
quark sources at nonzero temperature was discussed.

In the present paper, we continue to investigate lattice QCD as
initiated in \cite{KR}. Our starting point is the notion of the
algebra ${\mathfrak A}_{\Lambda}$ of field operators, see Section
\ref{FieldAlgebra}. After discussing local gauge transformations,
the Gauss law and boundary data, we show uniqueness of irreducible
representations of ${\mathfrak A}_{\Lambda}$. In Section
\ref{observablealgebra}, the algebra of internal observables
${\mathfrak O}^i_{\Lambda}$ is defined as the algebra of gauge
invariant fields, satisfying the Gauss law. We show that its
irreducible representations are labeled by distributions of colour
electric fluxes running through the boundary to infinity. It is
remarkable that for the very classification of irreducible
representations the abstract characterization of ${\mathfrak
O}^i_{\Lambda}$ as the subalgebra, invariant under the group of
local gauge transformations, factorized with respect to the ideal
generated by the Gauss law, is sufficient. This is due to the fact
that here we work within the compact formulation, which implies
that we ``stay within'' the representation space of the field
algebra. This remark does not apply to QED in the non-compact
formulation.

In chapter 4, we explain that in order to implement a strategy to
construct the thermodynamical limit {\em via} an inductive
(resp.~projective) limit procedure for observable algebras
(resp.~state spaces), we have to extend both the algebra of
internal observables ${\mathfrak O}^i_{\Lambda}$ (by adding
certain ``external observables'') and the Hilbert space ${\cal
H}_{\Lambda}$ (by tensorizing with the Hilbert space of tensors at
infinity). These external observables enable us to take into
account the correlations between the field degrees of freedom
contained in $\Lambda$ and the ``rest of the world''. The algebra
obtained this way is called full observable algebra and is denoted
by ${\mathfrak O}_{\Lambda}$. Within this context, we show that
different Hilbert space sectors, corresponding to different
boundary flux distributions, carry equivalent representations of
${\mathfrak O}_{\Lambda}$ if and only if their global ${\mathbb
Z}_3$-valued flux is the same, with intertwiners given by charge
carrying fields. This reduces the irreducible representations of
${\mathfrak O}_{\Lambda}$ to three inequivalent sectors, labeled
by the ${\mathbb Z}_3$-valued global colour electric flux. By the
global Gauss law, this flux coincides with the global colour
charge (triality) carried by the quark fields.

Finally, in Section \ref{Observablealgebra}, ${\mathfrak
O}_{\Lambda}$ is discussed in terms of generators and relations.
We start with presenting a set of genuine invariants generating
${\mathfrak O}_{\Lambda}$, which is, however, highly redundant. In
the remainder of this section, we use some gauge fixing methods to
reduce this set. In this context, a couple of delicate questions
arises -- all in some sense related to the Gribov problem and to
the fact that the underlying classical configuration space has a
complicated stratified structure with respect to the gauge group
action. A more complete treatment of ${\mathfrak O}_{\Lambda}$ as
an algebra presented by generators and relations will be given in
two separate papers, see \cite{CKR} and \cite{JKR}.

Finally, we mention that we have made some attempts to formulate
continuum gauge models in terms of observables within the
functional integral approach, see \cite{KRR}, \cite{KRR1} and
further references therein.


\setcounter{equation}{0}
\section{The Field Algebra}
\label{FieldAlgebra}


\subsection{Basic definitions}
\label{defineFieldAlgebra}


We consider QCD in the Hamiltonian framework on a finite regular
cubic lattice $\Lambda \subset \mathbb{Z}^3$, with  $\mathbb{Z}^3$
being the infinite regular lattice in 3 dimensions. We denote the
lattice boundary by $\partial \Lambda$ and the set of oriented,
$i$-dimensional elements of $\Lambda$, respecitively $\partial
\Lambda$, by ${\Lambda}^i$, respectively $\partial \Lambda^i$,
where $i = 0,1,2,3 \, .$ Such elements are (in increasing order of
i) called sites, links, plaquettes and cubes. Moreover, we denote
the set of external links connecting boundary sites of $\Lambda$
with ``the rest of the world'' by $\Lambda^1_{\infty}$ and the set
of endpoints of external links at infinity by
$\Lambda^0_{\infty}$. For the purposes of this paper, we may
assume that for each boundary site there is exactly one link with
infinity. Then, external links are labeled by boundary sites and
we can denote them by $(x,\infty )$ with $x \in \partial
\Lambda^0$. The set of non-oriented $i$-dimensional elements will
be denoted by $|{\Lambda}|^i$. If, for instance, $(x,y) \in
{\Lambda}^1$  is an oriented link, then by $|(x,y)| \in
|{\Lambda}|^1$ we mean the corresponding non-oriented link. The
same notation applies to $\partial \Lambda^1$ and
$\Lambda_{\infty}^1$.

The basic fields of lattice QCD are quarks living at lattice sites
and gluons living on links, including links connecting the lattice
under consideration with ``infinity''. The field algebra is thus,
by definition, the $C^*$-tensor product of fermionic and bosonic
algebras:
\begin{equation}
\label{fieldalgebra}
{\mathfrak A}_{\Lambda} :=
{\mathfrak F}_{\Lambda} \otimes {\mathfrak B}_{\Lambda} \, ,
\end{equation}
with
\begin{equation}
\label{fermifieldalgebra} {\mathfrak F}_{\Lambda} := \bigotimes_{x
\in \Lambda ^0} {\mathfrak F}_x
\end{equation}
and
\begin{equation}
\label{bosonfieldalgebra}
{\mathfrak B}_{\Lambda} := {\mathfrak B}_{\Lambda}^i \otimes
{\mathfrak B}_{\Lambda}^b =
 \bigotimes_{|(x,y)| \in |\Lambda |^1} {\mathfrak B}_{|(x,y)|}
\bigotimes_{x \in \partial \Lambda^0} {\mathfrak B}_{|(x,\infty)|} \, .
\end{equation}
Here, ${\mathfrak B}_{\Lambda}^i$ and ${\mathfrak B}_{\Lambda}^b$
are the internal and boundary bosonic algebras respectively. We
impose {\em locality} of the lattice quantum fields by {\em
postulating} that the algebras corresponding to different elements
of $\Lambda$ (anti-)commute with each other.

\vspace{0.1cm}
\noindent
{\bf Remark:}\\
The bosonic boundary data represent nontrivial colour electric
flux through the boundary, which -- as will be seen later --
allows for non-trivial colour charge. As will be shown, nontrivial
boundary flux is necessary for taking into account correlations of
field degrees of freedom inside $\Lambda$ with the ``rest of the
world''. Even if, for some reasons, the global charge of the
Universe vanishes, there is no reason to assume that an arbitrary
finite part $\Lambda$ is also neutral.

\vspace{0.1cm}

The fermionic field algebra ${\mathfrak F}_x$ associated with a
lattice site $x$ is the algebra of canonical anticommutation
relations (CAR) of quarks at $x$. The quark field generators are
denoted by
\begin{equation}
{\Lambda}^0 \ni x \rightarrow  {\psi}^{aA}(x) \in {\mathfrak F}_x
\, ,
\end{equation}
where $a$ stands for bispinorial and (possibly) flavour degrees of
freedom and $A = 1,2,3$ is the colour index corresponding to the
fundamental representation of the gauge group $G = SU(3)$. (In
what follows, writing $G$ we have in mind $SU(3)$, but essentially
our discussion can be extended to arbitrary compact groups and
their representations.) The conjugate quark field is denoted by
${\psi^{*}}_{aA}(x)  \, ,$ where we raise and lower indices by the
help of the canonical hermitian metric tensor $g_{AB}$ in $\mathbb
C^3$ and the canonical skew-symmetric structure $\epsilon_{ab}$ in
the spinor space. The only nontrivial canonical anti-commutation
relations for generators of ${\mathfrak F}_x$ read:
\begin{equation}
\label{CCR3} [{\psi^{*}}_{aA}(x), \psi^{bB}(x)]_+ = {\delta^B}_A
{\delta^b}_a \, .
\end{equation}

The bosonic field algebra ${\mathfrak B}_{|(x,y)|}$ associated
with the non-oriented link $|(x,y)|$, (where $y$ also stands for
$\infty$), is given in terms of its isomorphic copies ${\mathfrak
B}_{(x,y)}$ and ${\mathfrak B}_{(y,x)}$, corresponding to the two
orientations of the link $(x,y)\, .$ The algebra ${\mathfrak
B}_{(x,y)}$ is generated by matrix elements of the gluonic gauge
potential on the link $(x,y) \, ,$
\begin{equation}
{\Lambda}^1 \ni (x,y) \rightarrow  {U^A}_B(x,y) \in  {\mathfrak
C}_{(x,y)} \, ,
\end{equation}
with ${\mathfrak C}_{(x,y)} \cong C(G)$ being the commutative
$C^*$-algebra of continuous functions on $G$ and $A,B = 1,2,3$
denoting colour indices, and by colour electric fields, spanning
the Lie algebra ${\mathfrak g}_{(x,y)} \cong  su(3)$. Choosing an
orthonormal basis $\{t_i \} \, ,$ $i = 1, \dots , 8$, of $su(3)$
we denote by $\{ E_i(x,y)\}$ the corresponding basis of
${\mathfrak g}_{(x,y)}$,
\begin{equation}
{\Lambda}^1 \ni (x,y) \rightarrow  {E}_i(x,y) :=  t_i \in
{\mathfrak g}_{(x,y)} \, .
\end{equation}
These elements generate, in the sense of Woronowicz \cite{unb},
the $C^*$-algebra ${\mathfrak P}_{(x,y)} \cong C^*(G)$, see
\cite{Ped,Ki} for a definition of the group $C^*$-algebra $C^*(G)$.

Observe that $G$ acts on $C(G)$ naturally by the left regular
representation,
\begin{equation}
\label{left} L_g(u)(g') := u(g^{-1} g') \, \, , \, \,
u \in C(G) \, .
\end{equation}
Differentiating this relation, we get an action of $e \in su(3)$
on $u \in C^{\infty}(G)$ by the corresponding right invariant
vector field $e^R$. Thus, we have a natural commutator between
generators of ${\mathfrak P}_{(x,y)}$ and smooth elements of
${\mathfrak C}_{(x,y)}:$
\begin{equation}
\label{commut-e-f} i \ [e , u ] :=  e^R(u)  \ .
\end{equation}

To summarize, we have a $C^*$-dynamical system $\left({\mathfrak
C}_{(x,y)},G,\alpha\right),$ with automorphism $\alpha$ given by
the left action (\ref{left}). The field algebra ${\mathfrak
B}_{(x,y)}$ is, by definition, the corresponding crossed product
$C^*$-algebra,
\begin{equation}
\label{balgebra-xy} {\mathfrak B}_{(x,y)} :=
 {\mathfrak C}_{(x,y)} \otimes_{\alpha} G \, ,
\end{equation}
see \cite{Ped,Brat} for these notions.

\vspace{0.2cm}
\noindent {\bf Remarks:}
\begin{enumerate}
\item The crossed product algebra $C(G) \otimes_{\alpha} G$ is a
$C^*$-algebra without unit, defined as the completion of
$L^1(G,C(G))$ in the sup-norm taken over all Hilbert space
representations of $L^1(G,C(G)) \, .$ It can be viewed as a skew
tensor product of $C(G)$ with $C^*(G)$ in the following sense: For
each $u \in C(G)$ and $f \in L^1(G)$ denote by $u \otimes f$ the
element of $L^1(G,C(G))$ given by $(u \otimes f)(g) := u f(g) \,
.$ Then, the linear span of such elements, with $f$ taken from a
dense subset of $L^1(G)$, is dense in $L^1(G,C(G)) \, .$ Note that
$C^*(G) \subset C(G) \otimes_{\alpha} G$, with the canonical
injection given by $$ f \rightarrow 1 \otimes f \, . $$ As already
noted, the group algebra $C^*(G)$ is a $C^*$-algebra generated by
unbounded elements in the sense of Woronowicz, see \cite{unb}.
Consequently, $C(G) \otimes_{\alpha} G$ is of this type, too. We
stress that the $su(3)$-generators $e$ of $C^*(G)$ do not belong
to the algebra, but are only affiliated in the $C^*$-sense. \item
Formula (\ref{commut-e-f}) is a natural generalization of the
Heisenberg commutation relation $[p,q]= - i$, describing the
abelian case $G = \mathbb{R}^1$. It corresponds to ``canonical
quantization'' over the phase space
$$
  T^*(G) \cong \mathfrak{g}^* \times G \ ,
$$
with $\mathfrak{g} = su(3)$ being the Lie algebra of $G$ and
$\mathfrak{g}^*$ being the dual space. Quantization applies to
functions on $\mathfrak{g}^* \times G$, hence, in particular, to
elements of $\mathfrak{g}$ . Thus, from the purely algebraic point
of view, one would then define ${\mathfrak P}_{(x,y)}$ as the
enveloping algebra ${\mathfrak U}({\mathfrak g})$ of
$\mathfrak{g}$, see \cite{KR}, yielding for the bosonic field
algebra ${\mathfrak B}_{(x,y)}$ the following crossed product
structure of Hopf algebras: $$ C^{\infty}(G) \otimes_{\alpha}
{\mathfrak U}({\mathfrak g}) \,. $$ This is an example of a
Heisenberg double of Hopf algebras, see of \cite{Schm,Kash}. This
choice has, however, substantial drawbacks related to the fact
that the operators assigned to the Lie algebra elements $e$ are
necessarily unbounded. This is why we choose the
functional-analytic framework above, where all observables are
{\em bounded} operators in a Hilbert space. We call the algebra
(\ref{balgebra-xy}) the  algebra of {\em generalized canonical
commutation relations} (CCR) over the group $G$. \item Within the
above framework, one can prove a generalization of the classical
uniquess theorem by von Neumann \cite{JvN}, stating that any
irreducible representation of the above CCR-algebra is equivalent
to the {\em generalized Schr\"odinger representation}, acting on
the Hilbert space $L^2(G)$ (with respect to the Haar measure).
This will be shown in Subsection \ref{fieldirreps}.
\end{enumerate}

The transformation law of elements of ${\mathfrak B}_{(x,y)}$
under the change of the link orientation is derived from the fact
that the (classical) $G$-valued parallel transporter $g(x,y)$ on
$(x,y)$ transforms to $g^{-1}(x,y)$ under the change of
orientation. This transformation lifts naturally to an isomorphism
\begin{equation}
\label{izo-I} {\cal I}_{(x,y)} : {\mathfrak B}_{(x,y)} \rightarrow
{\mathfrak B}_{(y,x)} \,
\end{equation}
of field algebras, defined by:
\begin{equation}\label{1izo}
{\cal I}_{(x,y)}(f):= \breve{f} \  \ \  , \ \ \
{\cal I}_{(x,y)}(e):= \breve{e} \, ,
\end{equation}
where $\breve{f}(g):= f(g^{-1})$ and $\breve{e}$ is the left
invariant vector field on $G$, generated by $-e$. The bosonic
field algebra ${\mathfrak B}_{|(x,y)|}$ is obtained from
${\mathfrak B}_{(x,y)}$ and ${\mathfrak B}_{(y,x)}$ by identifying
them via ${\cal I}_{(x,y)}$.

Now, we give a full list of relations satisfied by generators of
${\mathfrak B}_{|(x,y)|}$. Being entries of the fundamental
representation of $SU(3)\, ,$ the generators of ${\mathfrak
C}_{(x,y)}$ have to fulfil the following conditions:
\begin{eqnarray}
\label{unitary1} ({U^A}_B(x,y))^* {U_A}^C(x,y) & = & {\delta^C}_B
\ {\bf 1} \ ,
\\
\epsilon_{ABC}\  {U^A}_D(x,y) {U^B}_E(x,y) {U^C}_F(x,y) & = & \
\epsilon_{DEF} \ {\bf 1} \ . \label{unitary2}
\end{eqnarray}
In what follows, we will use the traceless matrix
\begin{equation}
{E^A}_B(x,y) := \sum_i E_i (x,y) { { t_i}^A}_B \  , \label{EABEi}
\end{equation}
built from generators of ${\mathfrak P}_{(x,y)}$. Its entries
obviously fulfil
\begin{eqnarray}
\label{self-adj} ({E^A}_B(x,y))^* & = & {E_B}^A(x,y) \, .
\end{eqnarray}
The transformation law (\ref{1izo}) of these objects under the
change of the link orientation is given by the following
relations:
\begin{eqnarray}\label{izo-U}
  {U^A}_B(y,x) &=& {{\breve U}^A}_{\ \ B}(x,y) = ({U_B}^A(x,y))^* \ ,
  \\
  \label{ER-EL-1}
  {E^A}_B(y,x)  &=&  {{\breve E}^A}_{\ \ B}(x,y) =
   - {U^A}_D(y,x) {U^C}_B(x,y) {E^D}_C(x,y)  \ .
\end{eqnarray}

The $su(3)$-commutation relations read
\begin{equation}
\label{CCR2} [{E^A}_B(x,y) ,{E^C}_D(u,z)]  =  \delta_{xu}
\delta_{yz} \left({\delta^C}_B {E^A}_D(x,y)  - {\delta^A}_D
{E^C}_B(x,y) \right)  \ ,
\end{equation}
(formula (\ref{1izo}) implies that all the components
${E^A}_B(x,y)$ commute with all the components ${E^C}_D(y,x)$,
because the left invariant and the right invariant fields on the
group commute). The canonical commutation relations
(\ref{commut-e-f}) take the following form:
\begin{eqnarray}
\label{CCR1} i \ [{E^A}_B(x,y),{U^C}_D(u,z)] & = & + \delta_{xu}
\delta_{yz} \left({\delta^C}_B {U^A}_D(x,y) -\frac{1}{3}
{\delta^A}_B {U^C}_D(x,y) \right) \nonumber \\ & & - \delta_{xz}
\delta_{yu} \left({\delta^A}_D {U^C}_B(y,x) -\frac{1}{3}
{\delta^A}_B {U^C}_D(y,x) \right)  \, .
\end{eqnarray}

\vspace{0.2cm}
\noindent
To summarize, the field algebra ${\mathfrak A}_{\Lambda}$, given by
(\ref{fieldalgebra}) --  (\ref{bosonfieldalgebra}),
is a $C^*$-algebra, generated by elements
\begin{equation}
\label{gen-F}
  \left\{ \psi^{aA}(x) \, , \, {\psi^{*}}_{aA}(x)
  \, , \, {U^A}_B(x,y) \, , \, {E^A}_B(x,y)
  \right\} \, ,
\end{equation}
fulfilling relations (\ref{unitary1}), (\ref{unitary2}),
(\ref{self-adj}), (\ref{izo-U}) and (\ref{ER-EL-1}), together with
canonical (anti-) commutation relations (\ref{CCR3}), (\ref{CCR2})
and (\ref{CCR1}).

\noindent
{\bf Remark:} According to formula (\ref{ER-EL-1}), the transformation
of the colour electric field $E$ from $(x,y)$ to $(y,x)$ consists
of two steps: \\
1) the parallel transport from point $x$ to $y$
by means of the two parallel transporters $U$, each of them acting
appropriately on the two indices of $E$ \\ 
2) the change of the sign due to the change of the orientation. \\
In what follows, we always treat $E(x,y)$ as being
attached to the site $x$.


\subsection{Uniqueness of irreducible representations}
\label{fieldirreps}


Here, we prove uniqueness of the generalized canonical commutation
relations as announced in Subsection \ref{defineFieldAlgebra}.

We use the one-one-correspondence between non-degenerate
representations of crossed products and covariant representations
of $C^*$-dynamical systems, see \cite {Ped}. Thus, let us consider
the following covariant representation of $(C(G),G,\alpha)$ on
$L^2(G) \, ,$ (with respect to the Haar measure):
\begin{enumerate}
\item
Take the representation $\pi$ of the commutative $C^*$-algebra
$C(G)$ given by multiplication with elements of $C(G) \, .$ This
is, obviously, a representation by bounded operators on $L^2(G) \,
.$
\item
Consider the left regular (unitary) representation $\hat \pi$ of
$G$ on $L^2(G) \, ,$
\begin{equation}
\label{leftreg} ({\hat \pi}(g)h)(g') := h(g^{-1} g') \, \, , \, \,
h \in L^2(G) \, .
\end{equation}
We calculate
\begin{eqnarray}
 \left({\hat \pi}(g) \circ \pi(u) \circ
 {\hat \pi}(g^{-1})\right)(h)(g')
 & & =  \left(\left(\pi(u) \circ {\hat \pi}(g^{-1})\right)(h)\right)
 (g^{-1} g') \nonumber
 \\
 & & =  u(g^{-1} g') \cdot \left({\hat \pi}(g^{-1})(h)\right)
 (g^{-1} g') \nonumber
 \\
 & & =  u(g^{-1} g') \cdot h(g') \nonumber
 \\
 & & =  \pi \left( \alpha_g(u) \right)(h)(g') \, .\nonumber
\end{eqnarray}
This yields
\begin{equation}
\label{commpi-pi-hat1}
{\hat \pi}(g) \circ \pi(u) \circ {\hat
\pi}(g^{-1}) =  \pi \left( \alpha_g(u) \right) \, ,
\end{equation}
showing that the pair $(\pi, {\hat \pi})$ defines a {\em
covariant} representation of $(C(G),G,\alpha)$ on $L^2(G) \, ,$
indeed.
\end{enumerate}

The corresponding non-degenerate representation of $C(G)
\otimes_{\alpha} G$ is called {\em left regular representation}.
In the physical context, we call it {\em generalized Schr\"odinger
representation}. Observe that differentiating relation
(\ref{commpi-pi-hat1}) yields the generalized canonical
commutation relations (\ref{commut-e-f}).

Now, consider an arbitrary {\em covariant} representation $(\rho,
\hat \rho)$ of the the $C^*$-dynamical system $(C(G),G,\alpha)$,
with $\rho$ being a nondegenerate representation of $C(G)$ on a
Hilbert space $H$ and $\hat \rho$ being a strongly continuous
unitary representation of $G$ on $H$. By the Gelfand-Najmark
theorem for commutative $C^*$-algebras, we have a spectral measure
$dE$ on $G \, ,$ such that
\begin{equation}
\label{pispec} \rho(u) =  \int u(g) \, dE(g) \, ,
\end{equation}
for $u \in C(G) \, ,$ and by covariance, the pair $(\rho,\hat
\rho)$ also fulfils (\ref{commpi-pi-hat1}). Thus, we get
\begin{equation}
\label{commpi-pi-hat2}
  {\hat \rho}(g) \circ  dE(g') \circ {\hat \rho}(g^{-1})
  =  dE(g g')  \, .
\end{equation}
We conclude that the spectral measure $dE$ defines a transitive
system of imprimitivity for the representation ${\hat \rho}$ of
$G$ based on the group manifold $G\, .$ Then, the imprimitivity
theorem, see \cite{BaRa},\cite{Ki1}, yields the following

\begin{Theorem}
\label{CCR} Any (non-degenerate) irreducible representation of
$C(G) \otimes_{\alpha} G$ is equivalent to the generalized
Schr\"odinger representation.
\end{Theorem}

\vspace{0.2cm} \noindent {\bf Remarks:}
\begin{enumerate}
\item Disregarding the $C^*$-context, Theorem \ref{CCR} is a
classical result of Mackey, see \cite{Ki1} and references therein.
\item Within the $C^*$-context, there is a formulation of the
commutation relations (\ref{commpi-pi-hat1}) for an arbitrary
locally compact group in terms of the pentagon equation, which
generalizes to quantum groups \cite{SLW1}.
\end{enumerate}

The following statement is a simple consequence of Theorem 7.7.12
in \cite{Ped}.
\begin{Lemma}
\label{compact} For any compact Lie group, the generalized
Schr\"odinger representation defines the following isomorphism of
$C^*$-algebras: $$ C(G) \otimes_{\alpha} G \cong
\mathfrak{K}(L^2(G)) \, , $$ where $\mathfrak{K}(L^2(G))$ denotes
the algebra of compact operators on $L^2(G)$.
\end{Lemma}

Now, we take the tensor product of generalized Schr\"odinger
representations over all links:
\begin{equation}
\label{productSchroedinger} \bigotimes_{(x,y) \in \Lambda^1}
L^2({\cal C}_{(x,y)}) \bigotimes_{x \in {\partial \Lambda}^0}
L^2({\cal C}_{(x,\infty)}) \cong L^2({\cal C}) \, , \,
\end{equation}
with
$$ {\cal C} :=  \prod_{(x,y) \in \Lambda^1} {\cal C}_{(x,y)}
\, \prod_{x \in {\partial \Lambda}^0} {\cal C}_{(x,\infty)}
$$
and each of the spaces ${\cal C}_{(x,y)}$ being diffeomorphic to
the group space of $G$.

This is, by Theorem \ref{CCR}, the unique representation space of
the gluonic field algebra ${\mathfrak B}_{\Lambda}$. Moreover,
using the classical uniqueness theorem for CAR-representations by
Jordan and Wigner \cite{JoWi}, any representation of fermionic
fields is equivalent to the fermionic Fock representation.
Consequently, using Lemma \ref{compact}, we get the following
\begin{Corollary}
\label{irrepslocalfa} The field algebra ${\mathfrak A}_{\Lambda}$
can be identified with the algebra ${\mathfrak K}(H_{\Lambda})$ of
compact operators on the Hilbert space
\begin{equation}
 \label{irreps}
  H_{\Lambda} =
  {\cal F}({\mathbb C}^{12N}) \otimes
  L^2({\cal C})
  \, ,
\end{equation}
with ${\cal F}({\mathbb C}^{12N})$ denoting the fermionic Fock
space generated by $12N$ anti-commuting pairs of quark fields.
\end{Corollary}

The subspace ${\cal F}({\mathbb C}^{12N})$ is spanned by vectors
\begin{equation}
\label{psi-s1-sn}
  {\psi^{*}}_{a_1 A_1}(x_1) \dots
  {\psi^{*}}_{a_n A_n}(x_n)
  |0> \, ,
\end{equation}
obtained from the fermionic Fock vacuum by the action of quark
creation operators. Consequently, any element of $H_{\Lambda}$ is
a linear combination of these {\em fermionic} vectors with
coefficients being $L^2$-functions depending on gluonic potentials
$U$.


\subsection{Gauge transformations and local Gauss law}
\label{Gauge--Gauss}


The group $G_{\Lambda}$ of local gauge transformations related to
the lattice ${\Lambda}$ consists of mappings
$$
\Lambda^0 \ni x \rightarrow g(x) \in G \, ,
$$
which represent internal gauge transformations, and of gauge
transformations at infinity,
$$
\Lambda^0_{\infty} \ni z \rightarrow g(z) \in G \, .
$$
Thus,
\begin{equation}
\label{gaugegroup} G_{\Lambda} := G_\Lambda^i \times
G_\Lambda^{\infty} = \prod_{x \in \Lambda^0} G_x \, \prod_{z \in
\Lambda^0_{\infty}} G_z \, ,
\end{equation}
with $ G_y \cong SU(3)\, , $ for every $y$. We denote the
corresponding Lie algebra by
\begin{equation}
  \label{gaugeLiealgebra}
  {\mathfrak g}_{\Lambda} := {\mathfrak g}_{\Lambda}^i
  \oplus {\mathfrak g}_{\Lambda}^{\infty}
  = \bigoplus_{x \in \Lambda^0} {\mathfrak g}_x \,
  \bigoplus_{z \in \Lambda^0_{\infty}} {\mathfrak g}_z
  \, ,
\end{equation}
with $ {\mathfrak g}_y \cong su(3)\, , $ for every $y$.

The group $G_{\Lambda}$ acts on the classical configuration space
${\cal C}$ as follows:
\begin{equation}\label{glu-gau}
   {\cal C}_{(x,y)} \ni g(x,y) \rightarrow g(x)
   g(x,y) g(y)^{-1} \in {\cal C}_{(x,y)} \, ,
\end{equation}
with  $g(x) \in G_x$ and $g(y) \in G_y$. This action lifts
naturally to functions on ${\cal C}$. Moreover, we have an action
of $G_x$ on itself by inner automorphisms. This yields an action
of $G_{\Lambda}$ by automorphisms on each $C^*$-dynamical system
$({\mathfrak C}_{(x,y)},G,\alpha)$ and, therefore, on the gluonic
field algebra ${\mathfrak B}_{\Lambda}$. For generators of
${\mathfrak B}_{(x,y)} \subset {\mathfrak B}_{\Lambda}$, this
action is given by
\begin{eqnarray}
\label{gluonic-gauge}
{U^A}_B(x,y) & \rightarrow & {g^A}_C(x)
{U^C}_D(x,y) {{(g^{-1})}^D}_B(y) \ ,
\\
{E^A}_B(x,y) & \rightarrow & {g^A}_C(x) {E^C}_D(x,y)
{{(g^{-1})}^D}_B(x) \ ,
\end{eqnarray}
with $y$ standing also for $\infty$.
Fermionic generators transform under the fundamental
representation:
\begin{equation}
\label{fermionic-gauge} \psi^{aA}(x)  \rightarrow  {g^A}_B(x)
\psi^{aB}(x) \ .
\end{equation}
To summarize, the group of local gauge transformation
$G_{\Lambda}$ acts on the field algebra ${\mathfrak A}_{\Lambda}$
in a natural way by automorphisms.

It is easy to check that, for $x \in \Lambda^0$, the above
automorphisms are generated by the following derivations of the
field algebra:
\begin{equation}
\label{calG}
  {{\cal G}^A}_B(x) := {\rho^A}_B(x) - \sum_{y\leftrightarrow x}
  {E^A}_B(x,y)  \ ,
\end{equation}
where ${y \leftrightarrow x}$ means that the sum is taken over all
nearest neighbours $y$ of $x$ (with $y$ also standing for
$\infty$), and where
\begin{equation}
\label{rho} {\rho^A}_B(x) = \sum_a \left( \psi^{*aA}(x)
{\psi^a}_{B}(x) - \frac 13 \delta ^A{}_B \psi^{*aC}(x)
{\psi^a}_{C}(x) \right) \
\end{equation}
is the local matter charge density, fulfilling ${\rho^A}_A(x) =
0$. Observe that both (\ref{calG}) and (\ref{rho}) satisfy the
$su(3)$-commutation relations separately and that the set $\left\{
{{\cal G}^A}_B(x) \right\}$ of generators spans the Lie algebra
${\mathfrak g}_{\Lambda}^i \, .$

The local Gauss law at $x \in {\Lambda}^0 $
reads
\begin{equation}
 \label{Gauss1}
\sum_{y \leftrightarrow x}
 {E^A}_B(x,y)  = {\rho^A}_B(x)\ ,
\end{equation}
meaning that the gauge generator ${{\cal G}^A}_B(x)$ defined by
formula (\ref{calG}) vanishes. Observe that for every $x \in
\partial \Lambda^0$, the corresponding boundary flux
$E^A{}_B(x,\infty)$ enters the Gauss law. All the Gauss laws at
boundary points can thus be easily ``solved'' by expressing the
boundary fluxes in terms of internal fields.

For $z \in \Lambda^0_\infty$, the generator (\ref{calG}) of gauge
transformations reduces to the boundary flux ${E^A}_B(z,x)$.
Non-vanishing of this flux means gauge dependence of the quantum
state under the action of $G_z \subset G_{\Lambda}^\infty$.
Neglecting these boundary fluxes means neglecting the possibility
that a non trivial colour charge occurs \cite{KR}. As will be
discussed in Subsection \ref{Motivation}, such a ``truncated
theory'' is not useful as a discrete approximation of the
continuum theory. (The continuum limit should be constructed
\cite{KRT} in terms of an inductive (resp.~projective) limit of
observable algebras (resp.~quantum states). In this context,
``external fluxes'' represent the necessary link between any two
intersecting lattices belonging to a whole sequence of lattices.)

\noindent
{\bf Remark:} We stress that the Gauss law (\ref{Gauss1}) is the
lattice counterpart of the ``covariant divergence law'' 
$$
D_k E^k \equiv \partial_k E^k + [A_k , E^k] = \rho 
$$
in the continuum theory. There, the volume integration yields
on the left hand side a standard boundary flux term 
(by applying Stokes theorem) and an additional volume integral contribution 
corresponding to the $[A_k , E^k]$-term. In our lattice formulation, the volume integration corresponds 
to summation over all local Gauss laws (\ref{Gauss1}). This yields 
a sum over boundary terms living on external links
$(x,\infty)$ {\em and} a volume contribution equal to $E(x,y) +
E(y,x)$ on each lattice link. This term mimics the term $[A_k ,
E^k]$ of the continuum theory, (e.~g.~it vanishes only if the parallel
transporter $U(x,y)$ is trivial, which corresponds to the case
$A_k=0$ in the continuum theory).


\setcounter{equation}{0}
\section{The Algebra of Internal Observables}
\label{observablealgebra}


\subsection{Basic definitions}
\label{defobsalgebra}


Physical observables, internal relative to $\Lambda$, are, by
definition, gauge invariant fields respecting the Gauss law.
Hence, we have to take the subalgebra
$$
  {\mathfrak A}^{G_{\Lambda}}
  \subset {\mathfrak A}_{\Lambda}
$$
of $G_{\Lambda}$-invariant elements of the field algebra
${\mathfrak A}_{\Lambda}$. This means, in particular, that
observables have to commute with all gauge generators $ {{\cal
G}^A}_B(x)$.

Moreover, we have to impose all relations inherited from the local
Gauss laws at all lattice sites ({\em not} including sites at
infinity) as defining relations of the observable algebra. This
means that the generators of $G_\Lambda^i$ have to vanish in all
possible gauge-invariant algebraic combinations. Hence, imposing
(\ref{Gauss1}) at the algebraic (representation-independent) level
means that we require vanishing of the ideal ${\mathfrak
I}^i_{\Lambda} \cap {\mathfrak A}^{G_{\Lambda}} $, with
${\mathfrak I}^i_{\Lambda}$ being the ideal generated by
${\mathfrak g}_{\Lambda}^i$. Thus, the algebra ${\mathfrak
O}^i_{\Lambda}$ of internal observables  is obtained from
${\mathfrak A}^{G_{\Lambda}} $ by factorizing with respect to this
ideal.
\begin{Definition}
\label{Def1} The algebra of internal observables relative to
$\Lambda$ is defined as
\begin{equation}
\label{O-equiv}
  {\mathfrak O}^i_{\Lambda} = {\mathfrak A}^{G_{\Lambda}} /
    \{{\mathfrak I}^i_{\Lambda} \cap
    {\mathfrak A}^{G_{\Lambda}} \} \, ,
\end{equation}
where ${\mathfrak A}^{G_{\Lambda}} \subset {\mathfrak
A}_{\Lambda}$ is the subalgebra of $G_{\Lambda}$-invariant
elements of ${\mathfrak A}_{\Lambda}$ and ${\mathfrak
I}^i_{\Lambda} \subset {\mathfrak A}_{\Lambda}$ is the ideal
generated by ${\mathfrak g}_{\Lambda}^i$.
\end{Definition}

\noindent {\bf Remarks :}\\ i)~The above ideal ${\mathfrak
I}^i_{\Lambda}$ is generated by unbounded elements in the sense of
Woronowicz. It is obtained by multiplying its generators (elements
of ${\mathfrak g}_{\Lambda}^i$) from both sides by elements of
${\mathfrak A}_{\Lambda}$ belonging to their common dense domain,
e.~g.~the so called {\em smooth} elements (corresponding to
$C^\infty$-functions on $G$).
\\ ii)\label{remarkii}~The notion of an {\em observable} in the
above sense is somewhat narrow, e.g. only compact operator
functions built from Wilson loops belong to ${\mathfrak
O}^i_{\Lambda}$. This is due to the fact that the discussion of
irreducible representations of the generalized commutation
relations for bosonic fields as discussed in Section
\ref{FieldAlgebra} necessarily leads to the algebra ${\mathfrak
K}(H_{\Lambda})$ of compact operators on $H_{\Lambda}$. Moreover,
it is only this algebra, for which the notion of ``being generated
by unbounded elements'' makes sense. If one gives up these basic
structures, one can extend the field algebra, for instance, to the
algebra of all bounded operators on $H_{\Lambda}$ (the multiplier
algebra of ${\mathfrak K}(H_{\Lambda})$) and take the commutant of
$G_{\Lambda}$ there.


\subsection{Classification of irreducible representations}
\label{irrepsobsalg}


By Corollary \ref{irrepslocalfa}, we can identify ${\mathfrak
A}_{\Lambda}$ with the algebra of compact operators acting on the
Hilbert space $H_{\Lambda}$, given by (\ref{irreps}). Under this
identification, we have a unitary representation of the gauge
group $G_\Lambda$ on $H_{\Lambda}$ and the subalgebra ${\mathfrak
A}^{G_{\Lambda}}$ can be viewed as the commutant $(G_{\Lambda})'$
of this representation in ${\mathfrak K}(H_{\Lambda})$.

Consider the closed subspace ${\cal H}_{\Lambda} \subset
H_{\Lambda}$ consisting of vectors, which are invariant with
respect to internal gauge transformations,
\begin{equation}
\label{Hinv}
  {\cal H}_{\Lambda} := \{ h \in H_{\Lambda} \, \, | \, \,
  G_\Lambda^i h = h \, \} \ .
\end{equation}

\begin{Theorem}
\label{structurobsalgebra} The algebra of internal observables is
canonically isomorphic with the algebra of those compact operators
on the Hilbert space ${\cal H}_{\Lambda}$, which commute with the
action of the group $G_\Lambda^{\infty}$,
\begin{equation}
  \label{O-calH}
  {\mathfrak O}^i_{\Lambda} \cong {\mathfrak K}({\cal
  H}_{\Lambda}) \cap    (G_\Lambda^{\infty})' \ .
\end{equation}
\end{Theorem}

\noindent {\bf Proof:} Consider the direct sum decomposition
\begin{equation}
  \label{decomp} H_{\Lambda} = {\cal H}_{\Lambda} \oplus
  {\cal H}_{\Lambda}^{\bot} \, ,
\end{equation}
with ${\cal H}_{\Lambda}^{\bot}$ denoting the orthogonal
complement of ${\cal H}_{\Lambda}$. Since the actions of
$G_\Lambda^i$ and $G_\Lambda^{\infty}$ commute, ${\cal
H}_{\Lambda}$ is invariant under the action of
$G_\Lambda^{\infty}$ and, thus, under the full gauge group
$G_{\Lambda}$. Consequently, by unitarity of $G_{\Lambda}$, ${\cal
H}_{\Lambda}^{\bot}$ is invariant, too. This implies the following
block-diagonal structure of elements of $G_{\Lambda}$ with respect
to the decomposition (\ref{decomp}):
$$ \left(\begin{array}{cc} A
& 0\\ 0 & B
\end{array}\right) \, \, , \,
$$ with $A$ and $B$ denoting unitary operators on ${\cal
H}_{\Lambda}$ and ${\cal H}_{\Lambda}^{\bot}$, respectively. It
can be easily shown that
\begin{equation}
\label{commutant}
(G_{\Lambda})' = \left\{ \left(\begin{array}{cc}
C & 0\\
0 & D
\end{array}\right) \in {\mathfrak K}(H_{\Lambda}) :
[A,C] = 0 = [B,D] \, , \, \mbox{for all}
\left(\begin{array}{cc}
A & 0\\
0 & B
\end{array}\right) \in G_{\Lambda} \right\}\, .
\end{equation}
Indeed, an arbitrary element
$
\left(\begin{array}{cc}
C & E\\
F & D
\end{array}\right)
$
belongs to $(G_{\Lambda})'$ iff
$$
AC = CA \, \, , \, \,  AE = EB \, \, ,\,\,
BF = FA \,\,  , \,\,  BD = DB \, ,
$$
for any
$
\left(\begin{array}{cc}
A & 0\\
0 & B
\end{array}\right) \in G_{\Lambda} \, .
$
For
$
\left(\begin{array}{c}
h\\
0
\end{array}\right) \in  {\cal H}_{\Lambda}
$
we have
$
\left(\begin{array}{c}
0\\
Fh
\end{array}\right) \in  {\cal H}_{\Lambda}^{\bot} \, .
$
On the other hand, any element of $G_\Lambda^i$ has the form
$
\left(\begin{array}{cc}
{\bf 1} & 0\\
0 & B
\end{array}\right) \, .
$
Thus,
$$
\left(\begin{array}{cc}
{\bf 1} & 0\\
0 & B
\end{array}\right)
\left(\begin{array}{c}
0\\
Fh
\end{array}\right) =
\left(\begin{array}{c}
0\\
BFh
\end{array}\right)
=
\left(\begin{array}{c}
0\\
Fh
\end{array}\right) \, ,
$$ for all elements of $G_\Lambda^i$, yielding
$
\left(\begin{array}{c}
0\\
Fh
\end{array}\right) \in {\cal H}_{\Lambda}\, .
$ Thus, $Fh = 0$, for all $h \in {\cal H}_{\Lambda} \, ,$ implying
$F = 0 \, .$ In an analogous way one shows $E = 0 \, .$ This gives
formula (\ref{commutant}).

We decompose
$
\left(\begin{array}{cc}
C & 0\\
0 & D
\end{array}\right) = \left(\begin{array}{cc}
C & 0\\
0 & 0
\end{array}\right) + \left(\begin{array}{cc}
0 & 0\\
0 & D
\end{array}\right) \, .
$ Since the restriction of a compact operator to a closed
subspace is compact, we have $ \left(\begin{array}{cc}
C & 0\\
0 & 0
\end{array}\right) \in {\mathfrak K}({\cal H}_{\Lambda}) \, .
$
Moreover,
$
\left(\begin{array}{cc}
0 & 0\\
0 & D
\end{array}\right) \in {\mathfrak I}^i_{\Lambda} \, .
$
This yields the direct sum decomposition
\begin{equation}
  \label{commutant1}
  (G_{\Lambda})' = \left({\mathfrak K}({\cal H}_{\Lambda})
  \cap (G_{\Lambda})' \right) \oplus \left({\mathfrak
  I}^i_{\Lambda} \cap (G_{\Lambda})' \right) \, .
\end{equation}
Consequently, the algebra of internal observables
$
  {\mathfrak O}^i_{\Lambda} = (G_{\Lambda})' /
    \{{\mathfrak I}^i_{\Lambda} \cap
    (G_{\Lambda})' \}
$
is represented by the direct sum complement
$$
  {\mathfrak K}({\cal H}_{\Lambda}) \cap (G_{\Lambda})'
  = {\mathfrak K}({\cal H}_{\Lambda}) \cap (G_\Lambda^i)'\cap
  (G_\Lambda^{\infty})'
$$
in $(G_{\Lambda})'$. Finally, by (\ref{commutant}) we have that
arbitrary elements of $(G_\Lambda^i)'$ have the form $
\left(\begin{array}{cc}
C & 0\\
0 & D
\end{array}\right) \, ,
$
with $C \in {\mathfrak K}({\cal H}_{\Lambda})$ and $[D,B] = 0 \,
,$ for any unitary $B$ acting on ${\cal H}_{\Lambda}^{\bot}$.
Thus, ${\mathfrak K}({\cal H}_{\Lambda}) \subset (G_\Lambda^i)' \,
,$ yielding the isomorphism (\ref{O-calH}). \qed
\\

The restriction of the unitary action of $G^{\infty}_{\Lambda}$ to
the subspace ${\cal H}_{\Lambda}$ is not irreducible. Thus, ${\cal
H}_{\Lambda}$ splits into the direct sum of irreducible subspaces.
Each irreducible representation of $G_z \, , z \in
\Lambda^0_{\infty} \, ,$ is labeled by its highest weight $(m,n)$
and is equivalent to the corresponding tensor representation in
the space ${\mathbb S}^{m}_{n}({\mathbb C}^3)$ of
$m$-contravariant, $n$-covariant, traceless and totally symmetric
tensors over ${\mathbb C}^3$, endowed with the natural scalar
product induced by the scalar product on ${\mathbb C}^3$.
Therefore, irreducible representations of $G^{\infty}_{\Lambda}$
are labeled by sequences of highest weights
\begin{equation}
  \label{highweight1}
  ({\bf m}, {\bf n}) = (m_{z_1}, \dots , m_{z_M}
  ; n_{z_1}, \dots , n_{z_M}) \ ,
\end{equation}
where $(z_1, \dots z_M)$ label the lattice sites at infinity.
These representations are equivalent to tensor products of
representations in spaces ${\mathbb
S}^{m_{z_i}}_{n_{z_i}}({\mathbb C}^3)$. Let us denote by ${\cal
H}^{({\bf m}, {\bf n})}_{\Lambda}$ the sum of all the irreducible
subspaces with respect to the action of $G^{\infty}_{\Lambda}$,
which carry the same type $({\bf m}, {\bf n})$. Then we have
\begin{equation}
\label{Hinvdecomp}
  {\cal H}_{\Lambda} = \bigoplus {\cal H}^{({\bf m},
  {\bf n})}_{\Lambda} \ .
\end{equation}
Obviously, every subspace of type $({\bf m}, {\bf n})$ is
invariant under the action of the observable algebra,
\[
  {\mathfrak O}^i_{\Lambda}
  {\cal H}^{({\bf m}, {\bf n})}_{\Lambda}
  \subset
  {\cal H}^{({\bf m}, {\bf n})}_{\Lambda} \ .
\]
This yields the following

\begin{Corollary}
The irreducible representations of ${\mathfrak O}^i_{\Lambda}$ are
labelled by highest weight representations $({\bf m}, {\bf n})$ of
$G_\Lambda^{\infty}$.  For any $({\bf m}, {\bf n})$, the
corresponding irreducible representation of ${\mathfrak
O}^i_{\Lambda}$ coincides with the algebra of those compact
operators on ${\cal H}_{\Lambda}^{({\bf m}, {\bf n})}$, which
commute with the action of the group $G_\Lambda^{\infty}$.
\end{Corollary}

We call the pair $({\bf m}, {\bf n})$ the boundary flux
distribution carried by the gluonic field. In the next subsection,
it will become obvious that all distributions $({\bf m}, {\bf n})$
occur.


\subsection{Explicit description of irreducible representations}


Now we give an explicit description of the above irreducible
representations, using the explicit form (\ref{irreps}) of
$H_{\Lambda}$. Any element of $H_{\Lambda}$ is a linear
combination of fermionic vectors (\ref{psi-s1-sn}) with
coefficients being $L^2$-functions depending on gluonic potentials
$U$. The invariant subspace ${\cal H}_{\Lambda} \subset
H_{\Lambda}$ is spanned by vectors from $H_{\Lambda}$, which are
scalars with respect to $G_\Lambda^i$. This means that for every
$x \in {\Lambda}^0$, all the colour indices $(A_1, \dots , A_n)$
of the fermionic state (\ref{psi-s1-sn}) must be saturated with
the upper indices of either $U^A{}_B(x,y)$ or the canonical tensor
$\epsilon^{ABC}$. After such contractions, we are -- in general --
left with free indices at infinity points $z_i \in
\Lambda^0_\infty$. Finally, such a vector can be multiplied  by
gauge invariant functions of gluonic potentials $U$. The general
form of such functions is as follows:
\[
   f([U]) = f( {\rm Tr} (U_{\gamma_1}), \dots ,  {\rm Tr}
   (U_{\gamma_n}) ) \ ,
\]
where $f$ is a function of $n$ scalar variables, each $\gamma =
(x_1,x_2, \dots ,x_{m})$ is an arbitrary closed lattice path and
$U_{\gamma}$ is the corresponding {\em parallel transporter} along
$\gamma$,
\begin{equation}
  \label{paralleltransporter} U_{\gamma}^A{}_B =
  U^A{}_{C_2}(x_1,x_2) \, U^{C_2}{}_{C_3}(x_2,x_3) \dots
  U^{C_{m-1}}{}_B(x_{m-1},x_{m}) \ .
\end{equation}
Let us denote the result of these
operations by
\begin{equation}\label{Psi-indeksy}
{\boldsymbol\Psi} =\left( \Psi^{\dots , A_1  \dots
   A_{m_{z_i}} , \dots}_{\dots , B_1  \dots  B_{n_{z_i}} , \dots}
   \right) \ .
\end{equation}
This is a collection of  $G_\Lambda^i$-invariant vectors belonging
to $H_{\Lambda}$, labeled by tensor indices assigned to boundary
points $z_i \in \Lambda^0_\infty$, i.~e.~an $H_{\Lambda}$-valued
tensor over
\[
  {\mathbb C}^{3 M} =
  \bigoplus_{z_i \in
  \Lambda^0_\infty} {\mathbb C}_{z_i}^3
  \, .
\]
Linear combinations of those elements span the invariant subspace
${\cal H}_{\Lambda}$, with coefficients built from products of
tensors
\begin{equation}\label{t}
      t(z_i) = \left( t^{ B_1  \dots
    B_{n_{z_i}}}_{A_1  \dots  A_{m_{z_i}}} (z_i)
    \right) \in {\mathbb T}^{n_{z_i}}_{m_{z_i}}({\mathbb C}^3)\ .
\end{equation}
The resulting vector belonging to ${\cal H}_{\Lambda}$ is a scalar
obtained by contraction of (\ref{Psi-indeksy}) with these tensors:
\begin{equation}
\label{Psi-final}
 \Psi   = \Psi ( t(z_1) , \dots , t(z_M) ) =
  t^{ B_1  \dots  B_{n_{z_i}}}_{A_1  \dots  A_{m_{z_i}}}(z_i)
   \dots
  \Psi^{\dots \, , A_1  \dots A_{m_{z_i}}  ,\,
  \dots}_{\dots \, , B_1  \dots  B_{n_{z_i}}  ,\, \dots }
    \ .
\end{equation}
Each of the irreducible components ${\cal H}_{\Lambda}^{({\bf m},
{\bf n})}$ is composed of combinations (\ref{Psi-final}), for
which all the tensors $t(z_i)$ are irreducible (symmetric,
traceless), i.~e.~where $t(z_i) \in {\mathbb
S}^{n_{z_i}}_{m_{z_i}}({\mathbb C}^3) \subset {\mathbb
T}^{n_{z_i}}_{m_{z_i}}({\mathbb C}^3)$. If $t$'s are not
irreducible, (\ref{Psi-final}) is a sum of irreducible components
belonging to different weights $({\bf m}, {\bf n})$, according to
the decomposition of tensors $t(z_i)$ into the sum of products of
irreducible (symmetric, traceless) tensors with canonical tensors
$\delta^A{}_B$, $\epsilon^{ABC}$ and $\epsilon_{ABC}$.


\setcounter{equation}{0}
\section{The Full Algebra of Observables}



\subsection{Motivation and basic definitions}
\label{Motivation}


One of the main perspectives of this work is the construction of
the thermodynamical limit of finite lattice QCD. In \cite{KRT} we
have, in the context of finite lattice QED, outlined a strategy
based upon an inductive (resp.~projective) limit procedure for
observable algebras (resp.~state spaces). In what follows, we will
argue that in order to implement this strategy, we have to extend
both the algebra of observables ${\mathfrak O}^i_{\Lambda}$ (by
adding certain ``external observables'') and the Hilbert space
${\cal H}_{\Lambda}$ (by tensorizing with the Hilbert space of
tensors at infinity). Then, each collection $\left( \Psi^{\dots ,
A_1 , \dots , A_{m_{z_i}} , \dots}_{\dots , B_1 , \dots ,
B_{n_{z_i}} , \dots}\right)$ of $G_\Lambda^i$-invariant
$H_{\Lambda}$-vectors labelled by free indices at infinity points
(see \ref{Psi-indeksy}) will constitute a physical state.

Thus, let us consider two lattices $\Lambda_1$ and $\Lambda_2$
which are disjoint $(\Lambda_1 \cap \Lambda_2 = \emptyset )$ and
have a common wall such that their sum $\widetilde \Lambda =
\Lambda_1 \cup \Lambda_2$ is also a cubic lattice. If $x \in
\Lambda_1$ and $y \in \Lambda_2$ are adjacent points in
$\widetilde \Lambda$, then we identify their infinities. This
joint infinity $z$ may be visualised e.~g.~as the middle point of
the connecting link $(x,y)$. The parallel transporter on $(x,y)$
is defined by
\begin{equation}
\label{skladanie}
    U^A{}_{B}(x,y) := U^A{}_{C}(x,z) \,
    U^C{}_{B}(z,y)\ .
\end{equation}
Observables internal relative to $\Lambda_1$ (respectively
$\Lambda_2$) are built, among others, from parallel transporters
$U_\gamma$ along lattice paths $\gamma$, which are completely
contained in $\Lambda_1$ (respectively $\Lambda_2$). On the other
hand, there exist observables internal relative to
$\widetilde\Lambda$, built from parallel transporters along paths
crossing the set of joint infinity points. Such observables
describe correlations between phenomena occurring in the two
disjoint regions $\Lambda_1$ and $\Lambda_2$. As examples,
consider the $\widetilde\Lambda$-internal observables
$$
J^{a b}_{\gamma}(x,y) := {\psi^{*a}}_{A}(x) \, U^A_{\gamma \, B} \,
\psi^{bB}(y) \, ,
$$
with $\gamma$ being a path from $x \in \Lambda_1$ to $y \in
\Lambda_2$, or observables $$ U_{\gamma} := U^A_{\gamma \, A} \, ,
$$ with $\gamma$ being a closed path lying partially in
$\Lambda_1$ and partially in $\Lambda_2$. (These are operators
belonging to the multiplier algebra of the observable algebra
${\mathfrak O}^i_{\widetilde\Lambda}$ and, whence, may be called
{\em observables} in a wider sense only -- see Remark ii) on page
\pageref{remarkii}). According to the above mentioned inductive
limit procedure, the observable algebra related to
$\widetilde\Lambda$ should be constructed from observables related
to $\Lambda_1$ and $\Lambda_2$. But, in order to construct
observables related to $\widetilde\Lambda$ of the above type
(describing correlations), we have to admit fields ``having free
tensor indices at infinity'', like ${\psi^{*a}}_{A}(x) \,
U^A_{\gamma_1 \, C}(x,z)$ and $U^C_{\gamma_2 \, B}(z,y) \,
\psi^{bB}(y)$, with $z$ being a joint infinity point. Quantities
of this type are usually called ``charge carrying fields'', they
were first introduced by Mandelstam (see \cite{Mandel}).

Observe that charge carrying fields do not act neither on ${\cal
H}_{\Lambda_1}$ nor on ${\cal H}_{\Lambda_2} \, .$ They carry the
fundamental (respectively its contragredient) representation of
$SU(3)$ associated with the corresponding point at infinity. Thus,
we have to extend the Hilbert space ${\cal H}_{\Lambda}$ by
tensorising it with the Hilbert space  ${\mathbb T}_{\infty}$
generated by the fundamental and its contragredient
representations of $SU(3)$ associated with all points at infinity:
\begin{equation}
\label{TdLambda}
   {\mathbb T}_{\infty} :=
   \bigotimes_{z \in \Lambda^0_\infty} {\mathbb T}(z) \, \, \,  ,
\, \, \,  {\mathbb T}(z) :=
    \bigoplus_{(m,n)} {\mathbb T}^{m}_{n}(z) \  .
\end{equation}
Here, ${\mathbb T}^m_n(z)$ denotes the space of all -- not
necessarily irreducible -- $m$-contravariant, $n$-covariant
tensors over ${\mathbb C}_{z}^3$. This way, we are led to consider
the Hilbert space ${\mathbb T}_{\infty} \otimes {\cal H}_{\Lambda}
\, .$ The action of the gauge group $G_\Lambda^{\infty}$ extends
in a natural way from ${\cal H}_{\Lambda}$ to this tensor product:
\begin{equation}
\label{gauge-extend}
    T(g) \left( t \otimes \Psi \right) :=
    t \otimes  \left( g \cdot \Psi \right)\ \,   , \, \, \,
    t \otimes \Psi \in {\mathbb T}_{\infty} \otimes
    {\cal H}_{\Lambda} \, , \, \, \,
    g \in G_\Lambda^{\infty} \, \ .
\end{equation}
On the other hand, the natural action of $G_\Lambda^{\infty}$ on
${\mathbb T}_{\infty}$ can be also extended to this product:
\begin{equation}
  \label{gru-tensor}
    R(g) \left( t \otimes \Psi \right) :=
    \left( g \cdot t\right)  \otimes
    \Psi  \ .
\end{equation}
It is clear that elements of ${\mathbb T}_{\infty} \otimes {\cal
H}_{\Lambda}$ may be represented as ``wave functions with free
boundary indices'', i.~e.~objects of type (\ref{Psi-indeksy}).
Indeed, such tensors can be naturally viewed as anti-linear
mappings
\begin{equation}
\label{dual}
   {\mathbb T}_{\infty} \ni s \mapsto
   \left(t \otimes \Psi  \right) (s) := ( s | t)
   \Psi  \in  {\cal H}_{\Lambda}  \ ,
\end{equation}
where $( \cdot | \cdot )$ denotes the scalar product in ${\mathbb
T}_{\infty}$. Obviously, wave functions (\ref{Psi-indeksy}) are
mappings of this type, too.

The above extension of the Hilbert space is necessary if we want
to construct the thermodynamical limit of the theory by the above
mentioned projective limit procedure. According to this procedure,
physical states related to $\Lambda_1$ are obtained from physical
states related to $\widetilde\Lambda$ by applying a projection
operator $P_{\Lambda_1 , \widetilde\Lambda }$, which consists in
averaging over the degrees of freedom located in $\Lambda_2$. More
precisely, consider a wave function $\tilde \psi \in {\cal
H}_{\tilde \Lambda}\, , $ given by (\ref{Psi-final}), take the
corresponding projector $|\tilde \psi> <\tilde \psi| \, ,$ split
all parallel transporters on links joining $\Lambda_1$ and
$\Lambda_2$ according to (\ref{skladanie}), and integrate over all
degrees of freedom related to $\Lambda_2$ (including those located
on external links of $\Lambda_2$). The result is a mixed state
related to $\Lambda_1\, ,$ which can be represented as a mixture
of pure states, each of them being a ${\cal H}_{\Lambda_1}$-valued
tensor ${\boldsymbol \Psi}$ with respect to joint infinities of
$\Lambda_1$ and $\Lambda_2$. In other words, the averaging
procedure produces, in general, free indices on the common
boundary between $\Lambda_1$ and $\Lambda_2$. To be consistent, we
must admit such free indices from the very beginning.

>From the above discussion we see that tensors $t(z)$ occurring in
elements $t \otimes \Psi  \in {\mathbb T}_{\infty} \otimes {\cal
H}_{\Lambda}$ are not {\em a priori} given c-number quantities,
but quantum averages over external (i.~e.~contained in
$\Lambda_2$) degrees of freedom.

There is, however, an additional requirement, which we impose on
physical states of the system: the averaging procedure described
above should be compatible with gauge transformations. Assume that
we average a state $\tilde \psi \in {\cal H}_{\widetilde\Lambda}$
over $\Lambda_2$, with the parallel transporters on links joining
$\Lambda_1$ and $\Lambda_2$ split according to (\ref{skladanie}).
Any gauge transformation $g$ at a common infinity point $z$
between $\Lambda_1$ and $\Lambda_2$, acting on $\tilde \psi$ can
be either implemented by the action of $G_{\Lambda_1}^{\infty}$ or
$G_{\Lambda_2}^{\infty}$. After averaging over $\Lambda_2$, the
action of the gauge group $G_{\Lambda_1}^{\infty}$ is of course
still represented by $T(g)$, whereas the action of
$G_{\Lambda_2}^{\infty}$ reduces to $R(g)$ representing gauge
transformations  in ``the rest of the world''. But, compatibility
of averaging with gauging means that the result of this gauge
transformation should not depend upon its implementation. Hence,
we postulate
\begin{equation}
  \label{grupa-tensor}
    T(g) {\boldsymbol \Psi} = R(g) {\boldsymbol \Psi} \, .
\end{equation}
As a result of our discussion, we define the physical Hilbert
space as
\begin{equation}
   \label{Hext-mn} {\bf H}_{\Lambda} :=
   \left\{ {\boldsymbol \Psi} \in {\mathbb T}_{\infty}
   \otimes {\cal H}_{\Lambda} \ | \  T(g)
   {\boldsymbol \Psi} = R(g)  {\boldsymbol \Psi}\, , \ \
   \mbox{\rm for any} \ \ \   g \in
   G_\Lambda^{\infty}  \right\} \ ,
\end{equation}
with gauge transformation being, according to the above
discussion, represented by $T$. The property (\ref{grupa-tensor})
is obviously not shared by elements, which have partially
contracted indices at infinity or indices which do not come from
the bosonic wave functions $U(x,\infty)$. To illustrate this, we
consider the following examples:
\begin{equation}
\label{r-examples}
   \Psi^{\dots , A_1  \dots A_m , \dots} _{\dots , B_1  \dots  B_n ,
   \dots} \, r^{B_n}_{A_m}(z) \, \, \, \mbox{or} \, \, \, \Psi^{\dots
   , A_1  \dots A_m , \dots} _{\dots , B_1  \dots  B_n , \dots} \,
   r_{B_{n+1}}^{A_{m+1}}(z) \, , \, \, \, \mbox{with} \, \, \,
   r(z) \in {\mathbb T}^{1}_{1}(z) \, .
\end{equation}
Thus, (\ref{grupa-tensor}) is fulfilled precisely by elements of
type (\ref{Psi-indeksy}) having all indices free. Admitting
objects of type $r$, which would live at joint infinity points
$z$, would mean admitting additional degrees of freedom, relevant
for the description of physical states on the lattice $\widetilde
\Lambda = \Lambda_1 \cup \Lambda_2 \, .$ In that case, these joint
infinity points could not be removed from $\widetilde \Lambda\, .$

Using (\ref{TdLambda}), we have
\begin{equation}\label{rozklad}
  {\bf H}_{\Lambda} =
  \bigoplus_{({\bf m}, {\bf n}) }
  {\bf H}_{\Lambda}^{({\bf m}, {\bf n})}  \ .
\end{equation}
Here, ${\bf H}_{\Lambda}^{({\bf m}, {\bf n})}$ denotes the
intersection of ${\bf H}_{\Lambda}$ with ${\mathbb
T}_{\infty}^{({\bf m}, {\bf n})} \otimes {\cal H}_{\Lambda}$,
where
\begin{equation}
\label{TdLambda-mn}
   {\mathbb T}_{\infty}^{({\bf m}, {\bf n})} :=
   \bigotimes_{z \in \Lambda^0_\infty} {\mathbb T}^{m_z}_{n_z}(z)
\end{equation}
is the subspace of tensorial type $(m_z,n_z)$ at each $z \in
\Lambda^0_\infty \, .$ Note that, contrary to (\ref{Hinvdecomp}),
(\ref{rozklad}) is {\em not} a decomposition into irreducible
components.

Next, observe that the scalar products on $H_{\Lambda}$ and on
${\mathbb T}_{\infty}$ induce a natural scalar product on ${\bf
H}_{\Lambda}$. Using the representation (\ref{Psi-indeksy}), it is
given by:
\begin{eqnarray}
   ({\boldsymbol \Psi} | {\boldsymbol \Phi} )_{{\bf H}_{\Lambda}}
   & = & \left(
   \Psi^{\dots , A_1 , \dots ,
   A_{m_{z_i}} , \dots}_{\dots , B_1 , \dots , B_{n_{z_i}} ,
   \dots} \right| \left.
   \Phi^{\dots , C_1 , \dots ,
   C_{m_{z_i}} , \dots}_{\dots , D_1 , \dots , D_{n_{z_i}} ,
   \dots} \right)_{{H}_{\Lambda}} \times \nonumber \\
   \label{scalar-prod}
   & & \times
   \dots g_{A_1 C_1} \dots g_{A_{m_{z_i}} C_{m_{z_i}}} \dots
   g^{B_1 D_1}\dots g^{B_{m_{z_i}} B_{m_{z_i}}} \dots \ .
\end{eqnarray}
Tensors with different valences (i.~e.~having a different number
of indices) are, by definition, orthogonal.

We define the full algebra of observables related to $\Lambda$ as
the $C^*$-algebra of gauge invariant compact operators acting on
${\bf H}_{\Lambda}$,
\begin{equation}
\label{Observables-final}
  {\mathfrak O}_{\Lambda} :=
  {\mathfrak K}({\bf H}_{\Lambda}) \cap
  (G^{\infty}_{\Lambda})^\prime \ .
\end{equation}
Consider the algebra ${\mathfrak O}^{\infty}_{\Lambda}$ of those
compact operators acting on ${\mathbb T}_{\infty}$, which are
invariant with respect to the action of $G^{\infty}_{\Lambda}$. By
classical invariant theory, this algebra is generated (in the
sense of Woronowicz) by operations of tensorizing or contracting
with $SU(3)$-invariant tensors $\delta^A{}_B$, $\epsilon^{ABC}$
and $\epsilon_{ABC}$, and by projection operators $P^{({\bf m},
{\bf n})}$ onto ${\mathbb T}_{\infty}^{({\bf m}, {\bf n})} \subset
{\mathbb T}_{\infty}$.

\begin{Proposition}
The full observable algebra can be characterized as follows:
\begin{equation}
\label{O=OxO}
{\mathfrak O}_{\Lambda}  \cong {\mathfrak O}^i_{\Lambda}
\otimes {\mathfrak O}^{\infty}_{\Lambda} \, .
\end{equation}
\end{Proposition}

\noindent {\bf Proof:} Any compact gauge invariant operator $A$
acting on ${\bf H}_{\Lambda}$ can be extended to a gauge invariant
operator on the whole tensor product ${\mathbb T}_{\infty} \otimes
{\cal H}_{\Lambda}$, using contractions and tensor products with
boundary tensors $r$ in ${\mathbb T}_{\infty}$, see formula
(\ref{r-examples}). More precisely, we put:
\begin{equation}\label{extension}
    A(C(r \otimes {\boldsymbol \Psi})):=
    C(r \otimes A({\boldsymbol \Psi})) \ ,
\end{equation}
for any tensor $r \in {\mathbb T}_{\infty}$ and any contraction
operator $C$ (the result of the contraction on the right-hand-side
vanishes {\em by definition} if a corresponding index is missing
in $A {\boldsymbol \Psi}$). This way we have proved that
\begin{equation}\label{iden}
    {\mathfrak K}({\bf H}_{\Lambda}) \cap
  (G^{\infty}_{\Lambda})^\prime \cong {\mathfrak K}
  ({\mathbb T}_{\infty} \otimes {\cal H}_{\Lambda})
  \cap
  (G^{\infty}_{\Lambda})^\prime \ .
\end{equation}
But we have
\begin{equation}
   \label{O-product}
   {\mathfrak K}({\mathbb T}_{\infty} \otimes
   {\cal H}_{\Lambda}) \cong
   {\mathfrak K}({\mathbb T}_{\infty})
   \otimes {\mathfrak K}({\cal H}_{\Lambda}) \, ,
\end{equation}
and, consequently,
\begin{equation}\label{iden2}
    {\mathfrak K}({\bf H}_{\Lambda}) \cap
  (G^{\infty}_{\Lambda})^\prime \cong
   {\mathfrak K}({\mathbb T}_{\infty})
   \otimes \left( {\mathfrak K}({\cal H}_{\Lambda})
  \cap
  (G^{\infty}_{\Lambda})^\prime \right) \cong
  {\mathfrak K}({\mathbb T}_{\infty}) \otimes
  {\mathfrak O}^i_{\Lambda}
  \ .
\end{equation}
Taking again the intersection with $(G^{\infty}_{\Lambda})^\prime$
and implementing it by the representation $R$ yields the thesis.
\qed

By abuse of language, elements of ${\mathfrak
O}^{\infty}_{\Lambda}$ can be called ,,external observables''. We
also recall once again that, according to Remark ii) on page
\pageref{remarkii}, gauge invariant (not necessarily compact)
operators acting on ${\bf H}_{\Lambda}$ can be called
,,generalized observables'' or observables in a broader sense.

Adopting the point of view that $t$ occurring in $t \otimes \Psi$
represents the ``quantum averages over the external field degrees
of freedom'', one can argue that the only trace of the action of
external observables, which may be seen from $\Lambda$, are
external gauge invariant operators on ${\mathbb T}_{\infty}$.
Hence, formula (\ref{O=OxO}) could be taken as an axiomatic
definition of the full observable algebra. The results of the
following subsection show that this approach is equivalent to the
one used in the present section.


\subsection{Classification of irreducible representations}
\label{classification}


Obviously, ${\mathbb T}_{\infty}$ is not irreducible with respect
to the action of ${\mathfrak O}^{\infty}_{\Lambda}$. If $t(z) \in
{\mathbb T}^m_n(z)$, then its image under this action is a sum of
components belonging to ${\mathbb T}^k_l(z) \, ,$ with $(k,l)$
fulfilling
$$
(m - n)\ \mbox{\rm mod}\ 3 = (k - l)\
\mbox{\rm mod}\ 3\, .
$$
We see that the ${\mathbb Z}_3$-valued flux
\begin{equation}
\label{flux-loc}
  {\Phi}(z):=(m_z - n_z)\ \mbox{\rm mod}\ 3
\end{equation}
through each external link $(x,z)$, $z \in \Lambda^0_{\infty}$, is
conserved under the action of ${\mathfrak O}^{\infty}_{\Lambda}$.
Let us denote the sequence of ${\mathbb Z}_3$-valued fluxes
assigned to all boundary points by
\[
\boldsymbol\Phi := (\Phi(z_1) , \Phi(z_2), \dots ) \ .
\]
In what follows, we call $\boldsymbol\Phi$ boundary flux
distribution. Consequently, we define the subspace ${\bf
H}_{\Lambda}^{\boldsymbol\Phi} \subset {\bf H}_{\Lambda}$ as the
space spanned by those tensors (\ref{Psi-indeksy}) which fulfill
condition $\Phi(z_i)=(m_{z_i} - n_{z_i})\ \mbox{\rm mod}\ 3$. In
other words, we put
\begin{equation}
\label{H--Phi}
  {\bf H}_{\Lambda}^{\boldsymbol\Phi} :=
  \bigoplus_{m(z_i) - n(z_i)\ {\rm mod} \ 3 \, = \, \Phi(z_i) }
{\bf H}_{\Lambda}^{({\bf m}, {\bf n})} \, ,
\end{equation}
with ${\bf H}_{\Lambda}^{({\bf m}, {\bf n})}$ given by
decomposition (\ref{rozklad}). Obviously, these subspaces are
invariant under the action of the full observable algebra
${\mathfrak O}_{\Lambda}$ and ${\mathfrak O}_{\Lambda}$ acts
irreducibly on each of them. Moreover, the physical Hilbert space
${\bf H}_{\Lambda}$ splits into the direct sum of them:
\begin{equation}\label{H-t}
  {\bf H}_{\Lambda} = \bigoplus_{\boldsymbol\Phi}
  {\bf H}_{\Lambda}^{\boldsymbol\Phi} \ .
\end{equation}
We obviously have
\begin{Lemma}
\label{sectors-loc-triality} The spaces ${\bf
H}_{\Lambda}^{\boldsymbol\Phi}$ provide all the irreducible
representations of the algebra of observables ${\mathfrak
O}_{\Lambda}$.
\end{Lemma}
We denote the irreducible component of ${\mathfrak O}_{\Lambda}\, ,$
acting on  ${\bf H}_{\Lambda}^{\boldsymbol\Phi}\, ,$ by
${\mathfrak O}_{\Lambda}^{\boldsymbol\Phi}\, .$

Now, we define the global flux associated with a given boundary
flux distributions $\boldsymbol\Phi$ putting
\begin{equation}
\label{flux-glob}
  \Phi_{\partial\Lambda} := \left( \sum_{z \in \Lambda^0_\infty}
  \Phi(z) \right) \ \mbox{\rm mod}\ 3  \ .
\end{equation}
Let us denote the total number of gluonic and antigluonic flux
lines running through the boundary by
\begin{eqnarray}\label{M}
  m &:=& \sum_{z_i \in \Lambda^0_\infty} m(z_i) \ ,\\
  n &:=& \sum_{z_i \in \Lambda^0_\infty} n(z_i) \ . \label{N}
\end{eqnarray}
Then, we get
\begin{equation}
\label{flux-glob1}
\Phi_{\partial\Lambda} = (m-n) \, \, \mbox{\rm mod}\ 3 \, .
\end{equation}

\begin{Lemma}
\label{sectors-triality} The irreducible representations of
${\mathfrak O}_{\Lambda}$ in ${\bf H}_{\Lambda}^{\boldsymbol\Phi}$
and in ${\bf H}_{\Lambda}^{\boldsymbol\Phi^\prime}$ are unitarily
equivalent, if and only if ${\boldsymbol\Phi}$ and
${\boldsymbol\Phi^\prime}$ carry the same global flux,
$$
\Phi_{\partial\Lambda}=\Phi^\prime_{\partial\Lambda} \, .
$$

\end{Lemma}

\noindent {\bf Proof:} Suppose that we are given a pair
$(\boldsymbol\Phi , \boldsymbol\Phi^\prime )$ such that
$\Phi_{\partial\Lambda}=\Phi^\prime_{\partial\Lambda}$. Then,
${\bf H}_{\Lambda}^{\boldsymbol\Phi}$ is given by (\ref{H--Phi})
and, similarly,
\[
  {\bf H}_{\Lambda}^{\boldsymbol\Phi^\prime} =
  \bigoplus_{m^\prime(z_i) - n^\prime(z_i)\ {\rm mod} \ 3 \,
  = \, \Phi^\prime(z_i) }
{\bf H}_{\Lambda}^{({\bf m}^\prime, {\bf n}^\prime)} \, .
\]
For $\Phi_{\partial\Lambda}=\Phi^\prime_{\partial\Lambda} \, ,$
formula (\ref{flux-glob1}) implies that we can choose a pair of
labels $({\bf m}_0, {\bf n}_0)$ and $({\bf m}_0^\prime, {\bf
n}_0^\prime)$ such that $m_0=m_0^\prime$, $n_0=n_0^\prime$, Acting
with tensorial operators $\xi \in {\mathfrak
O}^{\infty}_{\Lambda}$ (tensorising and contracting with canonical
tensors $\delta^A{}_B$, $\epsilon^{ABC}$ and $\epsilon_{ABC}$) on
${\bf H}_{\Lambda}^{({\bf m}_0, {\bf n}_0)}$ (respectively ${\bf
H}_{\Lambda}^{({\bf m}_0^\prime, {\bf n}_0^\prime)}$), we may pass
to any other subspace ${\bf H}_{\Lambda}^{({\bf m}, {\bf n})}$ of
${\bf H}_{\Lambda}^{\boldsymbol\Phi}$ (respectively any other
subspace ${\bf H}_{\Lambda}^{({\bf m}^\prime, {\bf n}^\prime)}$ of
${\bf H}_{\Lambda}^{\boldsymbol\Phi^\prime}$). This way we
construct a bijection between the two sets $({\bf m}, {\bf n})$
and $({\bf m}^\prime, {\bf n}^\prime)$ corresponding to
${\boldsymbol\Phi}$ and ${\boldsymbol\Phi^\prime}$, preserving the
total number of gluonic and antigluonic lines,  $m=m^\prime$ and
$n=n^\prime$.

We shall construct an intertwining operator between
representations on ${\bf H}_{\Lambda}^{\boldsymbol\Phi}$ and ${\bf
H}_{\Lambda}^{\boldsymbol\Phi^\prime} \, .$ For this purpose, we
first define a sequence of  isometric isomorphisms of Hilbert
spaces,
\[
  {\bf U}_{({\bf m}^\prime,
  {\bf n}^\prime)({\bf m}, {\bf n})} :
  {\bf H}_{\Lambda}^{({\bf m}, {\bf n})} \rightarrow
  {\bf H}_{\Lambda}^{({\bf m}^\prime, {\bf n}^\prime)}
\]
corresponding to the above bijection, as follows. We choose two
finite families $\{ \beta_a \}$ and $\{ \gamma_b \}$ of lattice
paths in $\Lambda$, fulfilling the following conditions:
\begin{itemize}
  \item their starting points $x_a$ and $y_b$, together with their
  end points $z_a$ and $w_b$ belong to the boundary
  $\partial\Lambda$,
  \item for every $x \in \Lambda^0_\infty$ there are exactly
  $\Delta n(x):=n(x) -
  n^\prime (x)$ paths $\beta$ starting from $x$ if $\Delta n(x)
  >0$ and zero otherwise,
  \item for every $x \in \Lambda^0_\infty$ there are exactly
  $\Delta m(x):=m(x) -
  m^\prime (x)$ paths $\gamma$ ending at $x$ if $\Delta m(x)
  >0$ and zero otherwise,
  \item for every $x \in \Lambda^0_\infty$ there are exactly
  $- \Delta n(x)$ paths $\beta$ ending at $x$ if $\Delta n(x)
  < 0$ and zero otherwise,
  \item for every $x \in \Lambda^0_\infty$ there are exactly
  $- \Delta m(x)$ paths $\gamma$ starting from $x$ if $\Delta m(x)
  < 0$ and zero otherwise.
\end{itemize}
Now, the action of the operator ${\bf U}_{({\bf m}^\prime, {\bf
n}^\prime)({\bf m}, {\bf n})}$ on a vector ${\boldsymbol \Psi} \in
{\bf H}_{\Lambda}^{({\bf m}, {\bf n})}$ is defined as follows. We
multiply the tensor (\ref{Psi-indeksy}) by all parallel
transporters $U_{\beta_a}^{B_a}{}_{A_a}$ and
$U_{\gamma_b}^{B_b}{}_{A_b}$, see formula
(\ref{paralleltransporter}). Then, we contract all the subsequent
upper indices $B_a$ with the corresponding subsequent lower
indices $B_i$ of ${\boldsymbol \Psi}$ at the starting points of
the curves $\beta$ and all the subsequent lower indices $A_b$ with
the corresponding upper indices $A_i$ of ${\boldsymbol \Psi}$ at
the end points of the curves $\gamma$. It is easy to see  that the
inverse (adjoint) operator consists in multiplying ${\boldsymbol
\Psi}$ by the same transporters, but in contracting indices $A_a$
with the corresponding subsequent upper indices $A_i$ of
${\boldsymbol \Psi}$ at the starting points of the curves $\beta$
and all the subsequent indices $B_b$ with the corresponding lower
indices $B_i$ of ${\boldsymbol \Psi}$ at the end points of the
curves $\gamma$. This implies that
\[
 {\bf U}^*_{({\bf m}^\prime, {\bf n}^\prime)({\bf m}, {\bf n})}
 {\bf U}_{({\bf m}^\prime, {\bf n}^\prime)({\bf m}, {\bf n})}
 = \mbox{\rm id} =
 {\bf U}_{({\bf m}^\prime, {\bf n}^\prime)({\bf m}, {\bf n})}
 {\bf U}^*_{({\bf m}^\prime, {\bf n}^\prime)({\bf m}, {\bf n})} \
 .
\]
Organizing the operators ${\bf U}_{({\bf m}^\prime, {\bf
n}^\prime)({\bf m}, {\bf n})}$ into a block matrix, we get an
isometric isomorphism denoted by
\begin{equation}
\label{Utylda}
    {\bf U}_{{\boldsymbol\Phi^\prime} \boldsymbol\Phi}
    :{\bf H}_{\Lambda}^{\boldsymbol\Phi} \rightarrow
    {\bf H}_{\Lambda}^{\boldsymbol\Phi^\prime} \ .
\end{equation}

Next, observe that the action
$$
  {\bf U}_{{\boldsymbol\Phi^\prime} \boldsymbol\Phi}^* \
   {\mathfrak O}_{\Lambda}^{\boldsymbol\Phi^\prime} \
   {\bf U}_{{\boldsymbol\Phi^\prime} \boldsymbol\Phi}:
   {\bf H}_{\Lambda}^{\boldsymbol\Phi} \rightarrow
    {\bf H}_{\Lambda}^{\boldsymbol\Phi} \ ,
$$
defines an irreducible representation of ${\mathfrak O}_{\Lambda}$
on ${\bf H}_{\Lambda}^{\boldsymbol\Phi}$. By Lemma
\ref{sectors-loc-triality}, this representation must be unitarily
equivalent to ${\mathfrak O}_{\Lambda}^{\boldsymbol\Phi}$,
i.~e.~there exists a unitary (intertwining) operator
\begin{equation}
\label{S}
    {\bf S}
    :{\bf H}_{\Lambda}^{\boldsymbol\Phi} \rightarrow
    {\bf H}_{\Lambda}^{\boldsymbol\Phi} \ ,
\end{equation}
such that
$$
  {\bf S}^* {\bf U} _{{\boldsymbol\Phi^\prime} \boldsymbol\Phi}^* \,
  {\mathfrak O}_{\Lambda}^{\boldsymbol\Phi^\prime} \,
   {\bf U}_{{\boldsymbol\Phi^\prime} \boldsymbol\Phi}{\bf S}
   = {\mathfrak O}_{\Lambda}^{\boldsymbol\Phi} \ .
$$

It remains to show that equivalent representations yield equal
global fluxes: If ${\bf H}_{\Lambda}^{\boldsymbol\Phi}$ and ${\bf
H}_{\Lambda}^{\boldsymbol\Phi^\prime}$ carry equivalent
representations, then there exists an intertwiner ${\bf
V}_{{\boldsymbol \Phi^\prime}\boldsymbol\Phi}$, such that
$$
\Phi^\prime_{\partial\Lambda} = {\bf V}_{{\boldsymbol
\Phi^\prime}\boldsymbol\Phi}\, \Phi_{\partial\Lambda} \, {\bf
V}_{{\boldsymbol\Phi^\prime}\boldsymbol\Phi}^{-1} \,.
$$
But, according to formula (\ref{flux-glob1}),
$\Phi_{\partial\Lambda}$ is a scalar on every irreducible
representation space of ${\mathfrak O}_{\Lambda}\, . $ Thus, it
commutes with ${\bf V}_{{\boldsymbol \Phi^\prime}\boldsymbol\Phi}$
yielding $ \Phi^\prime_{\partial\Lambda} =  \Phi_{\partial\Lambda}
\,. $ \qed

\begin{Theorem}
\label{glob-flux} There are three inequivalent representations of
${\mathfrak O}_{\Lambda}$ labeled by values of the global flux
$\Phi_{\partial\Lambda}$. Consequently, the space ${\bf
H}_{\Lambda}$ splits into the sum of three eigenspaces of
$\Phi_{\partial\Lambda}$
\[
   {\bf H}_{\Lambda} = \bigoplus_{\lambda= -1,0,1}
   {\bf H}_{\Lambda}^\lambda \ .
\]
Each of the spaces ${\bf H}_{\Lambda}^\lambda$ is a sum of
superselection sectors ${\bf H}_{\Lambda}^{\boldsymbol\Phi}$
corresponding to all possible distributions ${\boldsymbol\Phi}$ of
the global flux $\lambda$. They carry equivalent representations
of ${\mathfrak O}_{\Lambda}$.
\end{Theorem}

Anticipating the final result, we call the spaces ${\bf
H}_{\Lambda}^\lambda$ ``charge superselection sectors'', in
contrast to the ``boundary-flux-distribution superselection
sectors'' ${\bf H}_{\Lambda}^{\boldsymbol\Phi}$. (For an analogous
discussion in continuum QED see \cite{Star}).


\subsection{Global colour charge and superselection structure}


Here we show that, according to the global Gauss law, irreducible
representations of ${\mathfrak O}_{\Lambda}$ are, alternatively,
labeled by global colour charge (triality), which is carried by
the quark fields. We briefly recall the notion of triality and
derive the global Gauss law, for details see \cite{KR}.

Consider any integrable representation $F$ of the Lie algebra
$su(3)$ on a Hilbert space $H \, ,$ i.~e.~a collection of
operators ${F^A}_B$ in $H$, fulfilling ${F^A}_A = 0 \, ,$
$({F^A}_B)^*  =  {F_B}^A$ and
\begin{equation}
\label{CCRF} [{F^A}_B , {F^C}_D] = {\delta^C}_B {F^A}_D  -
{\delta^A}_D {F^C}_B  \, .
\end{equation}
By (\ref{CCR2}), (\ref{rho}) and (\ref{CCR3}), the operators
${E^A}_B(x,y)$ and ${\rho^A}_B(x)$, occurring on both sides of the
local Gauss law, are of this type. Integrability means that for
each $F$ there exists a unitary representation $G \ni g
\rightarrow {\bar F}(g) \in B(H)$ of the group $G$ such that $F$
is its derivative. If $F_1$ and $F_2$ are two commuting
(integrable) representations of $su(3)$, then so is $F_1 + F_2$.
Such a collection of operators is an {\em operator domain} in the
sense of Woronowicz, see \cite{SLW}.

We define an operator function on this domain, i.~e.~a mapping $F
\rightarrow \varphi(F)\, ,$ which satisfies $\varphi(U F U^{-1} )
= U \varphi(F) U^{-1}$ for an arbitrary isometry $U$, as follows:
For any integrable representation $F$ of $su(3)$, consider the
corresponding representation ${\bar F}$ of $G$. Its restriction to
the center ${\cal Z}$ of $G$ acts as a multiple of the identity on
each irreducible subspace $H_\alpha$ of ${\bar F} \, ,$
$$
{\bar F}(z) |_{H_\alpha} = \chi^{\alpha}_{\bar F}(z)\cdot {\bf
1}_{H_\alpha} \, \, , \, \, \, z \in {\cal Z} \, . $$ Obviously,
$\chi^{\alpha}_{\bar F}$ is a character on ${\cal Z}$ and,
therefore, $(\chi^{\alpha}_{\bar F}(z))^3 = 1$. We identify the
group of characters on ${\cal Z} =\{ \zeta \cdot {\bf 1}_3 \ | \
\zeta^3 = 1 \ , \ \zeta \in {\mathbb C} \}$ with the additive
group ${\mathbb Z}_3 \cong \{-1,0,1\}$ by assigning to any
character $\chi^{\alpha}_{\bar F}$ a number $k(\alpha) \in \{ -1 ,
0 , 1 \}$ fulfilling
\[
  \chi^{\alpha}_{\bar
  F} (\zeta \cdot {\bf 1}_3) = \zeta^{k(\alpha)}  \, .
\]
Hence, there exists a  ${\mathbb Z}_3$-valued operator function
$F \rightarrow \varphi(F)$, defined by
\begin{eqnarray}
\label{innephi} \zeta^{\, \varphi_{\alpha} (F)} & = &
\chi^{\alpha}_{\bar F} (\, \zeta \cdot {\bf 1}_3) \, ,\\
\varphi(F) & = & \sum_{\alpha} \varphi_{\alpha} (F) \, {\bf
1}_{H_\alpha} \, . \label{full--phi}
\end{eqnarray}
Since $\chi^{\alpha}_{\bar F}$ are characters, we have
\begin{equation}
\label{fi3} \varphi(F_1 + F_2)  =   \varphi(F_1) + \varphi(F_2) \,
,
\end{equation}
for $F_1$ and $F_2$ commuting. Now, using the equivalence of each
irreducible representation $\alpha$ of $G$ with highest weight
$(m(\alpha),n(\alpha))$ with the tensor representation in the
space ${{\mathbb T}}^{m(\alpha)}{}_{n(\alpha)}({\mathbb C}^3)$ of
$m(\alpha)$-contravariant, $n(\alpha)$-covariant, completely
symmetric and traceless tensors over ${\mathbb C}^3$, we get
\begin{equation}
\label{innephi2} \chi^{\alpha}_{\bar F}(z) = \zeta^{\,
\varphi_{\alpha} (F)} = \zeta^{\, m(\alpha) - n(\alpha)}  \, ,
\end{equation}
for $z = \zeta \cdot {\bf 1}_3 \in {\cal Z} \, .$
Thus, we have
\begin{equation}
\label{varphi--m-n} \varphi_{\alpha} (F) = (m(\alpha)-n(\alpha))
\,\,\, \text{mod} \,\,\, 3
\end{equation}
for every irreducible highest weight representation
$(m(\alpha),n(\alpha)) \, .$ In \cite{KR} we have given an
explicit construction of $\varphi (F)$ in terms of Casimir
operators of $F \, .$

Applying $\varphi$ to the local Gauss law (\ref{Gauss1}) and using
additivity (\ref{fi3}) we obtain a gauge invariant equation for
operators with eigenvalues in ${\mathbb Z}_3$:
\begin{equation}
\label{inGauss}
\sum_{y\leftrightarrow x} \varphi(E(x,y)) =
\varphi(\rho(x)) \ ,
\end{equation}
valid at every lattice site $x$. The quantity on the right hand
side is the (gauge invariant) local colour charge density carried
by the quark field.

Using the transformation law (\ref{1izo}) for $E(x,y)$ under the
change of the link orientation and additivity (\ref{fi3}) of
$\varphi$, we have for every lattice bond $(x,y)$:
\begin{equation}
 \varphi(E(x,y)) + \varphi(E(y,x)) =
 \varphi(E(x,y)) + \varphi({\breve E}(x,y)) =
 \varphi(E(x,y) + {\breve E}(x,y))  \ ,
\end{equation}
because representations $E$ and $\breve E$ commute. But the
following identity holds:
\begin{equation}\label{inner}
  \varphi(E(x,y) + {\breve E}(x,y)) = 0 \ ,
\end{equation}
because both representations have the same irreducible subspaces
$H_\alpha$, with the values of  $m(\alpha)$ and $n(\alpha)$
exchanged. (The identity follows also directly from formula
(\ref{ER-EL-1}). It implies that the second order Casimirs for $E$
and $\breve E$ coincide: $K_2(E)=K_2(\breve E)$, whereas
$K_3(E)=-K_3(\breve E)$, cf. \cite{KR}).

Now, we take the sum of equations (\ref{inGauss}) over all lattice
sites $x \in \Lambda^0$. Due to the above identity, all terms on
the left hand side cancel, except for contributions
$\varphi(E(x,\infty )) \equiv \varphi(E(x,z))\, , \, z \in
\Lambda^0_{\infty}\, ,$ coming from the boundary. The irreducible
subspace ${\bf H}_{\Lambda}^{\boldsymbol\Phi} \subset {\bf
H}_{\Lambda} $ characterized by the flux distribution
${\boldsymbol\Phi}$, see (\ref{H--Phi}), is an eigenspace of the
gauge invariant operator $\varphi(E(x,z ))\, ,$ with eigenvalue
$(m_z - n_z)\ \mbox{\rm mod}\ 3 = {\Phi}(z) \, , $ according to
formula (\ref{varphi--m-n}). Thus, using (\ref{flux-glob}) we
obtain
\begin{equation}
\label{flux3}
  \sum_{x \in \partial \Lambda^0}
  \varphi(E(x,\infty )) = \Phi_{\partial \Lambda} \ ,
\end{equation}
with $\Phi_{\partial \Lambda}$ being the global ${\mathbb
Z}_3$-valued boundary flux corresponding to the flux distribution
${\boldsymbol\Phi}$. On the right hand side we get the (gauge
invariant) global colour charge ({\em triality}), carried by the
matter field
\begin{equation}
\label{globalM}
{\mathfrak{t}}_{\Lambda} := \sum_{x \in \Lambda^0}
\varphi(\rho(x)) \, .
\end{equation}
Thus, the {\em global Gauss law} takes the following form:
\begin{equation}\label{gG}
\Phi_{\partial \Lambda} = {\mathfrak{t}}_{\Lambda} \ .
\end{equation}
Both quantities appearing here take eigenvalues in the center
${\cal Z} \cong {\mathbb Z}_3$ of $G$.

Comparing with Theorem \ref{glob-flux}, the global Gauss law
yields another, equivalent characterization of irreducible
representations of the observable algebra:
\begin{Corollary}
\label{glob-triality} The inequivalent representations of
${\mathfrak O}_{\Lambda}$ are labeled by eigenvalues of global
colour charge ${\mathfrak{t}}_{\Lambda}$.
\end{Corollary}

\vspace{0.1cm}
\noindent
{\bf Remark:}\\
To illustrate the triality concept observe that we can assign to
single quark fields (being in the defining representation of
$SU(3)$) triality $+1$ and, consequently to antiquarks $-1$, (or
the other way around). A single lattice gluon field has, of
course, triality $0 \, .$ Now, imagine a state with $m$ quarks and
$n$ antiquarks, located in an arbitrary way inside $\Lambda\, .$
By additivity of triality, ${\mathfrak{t}}_{\Lambda}$ has
eigenvalue $m - n  \, \, \mbox{\rm mod}\ 3$ on this state. As
discussed already, gauge invariance of states with respect to
internal gauge transformations implies that all quark indices
inside $\Lambda$ have to be contracted. Basically, this can be
done by connecting quark-antiquark pairs {\em inside} $\Lambda$
with flux lines built from gluonic parallel transporters and by
contracting with canonical tensors $\delta^A{}_B$,
$\epsilon^{ABC}$ and $\epsilon_{ABC}$. On the other hand, some of
the flux lines starting at a quark (or ending at an antiquark)
inside can run through the boundary to end at an antiquark (or
start from a quark) {\em outside} of $\Lambda$. By the global
Gauss law, the total number of gluonic flux lines minus the number
of antigluonic flux lines, calculated modulo $3$, is equal to the
eigenvalue of triality. But the external quark and antiquark
fields are not taken into account by a theory on $\Lambda$. After
averaging over external fields we are left with the action of
external gauge invariant operators $\delta^A{}_B$,
$\epsilon^{ABC}$ and $\epsilon_{ABC}$. The flux lines running
through the boundary may by contracted at points $z\in
\Lambda^0_{\infty}$ with these tensors, eventually leaving either
none (${\mathfrak{t}}_{\Lambda}=0$), or one gluonic
(${\mathfrak{t}}_{\Lambda}=+1$) or one antigluonic
(${\mathfrak{t}}_{\Lambda}=-1$) non-contracted line.


\setcounter{equation}{0}
\section{Generators and Relations}
\label{Observablealgebra}


In this section we wish to find a presentation of the observable
algebra in terms of generators and relations, inherited from
canonical commutation relations of fields and from the Gauss law.
In order to formulate and to study field dynamics, we rather need
a presentation in terms of a set of independent generators. Thus,
we wish to solve the Gauss law relations explicitly, to end up
with a reduced set of generators and their (anti)-commutation
relations. We show how to implement this idea by a special gauge
fixing procedure based upon the choice of a lattice tree. This
procedure leaves us, however,  with some discrete gauge freedom.
Moreover, we restrict ourselves to the generic stratum of the
gauge group action, disregarding all non-generic strata. But even
there, our method works only on a dense subset. These two
obstructions to global gauge fixing reflect the Gribov problem,
which is well known in the continuum theory, see also
\cite{Fleisch} for a discussion of this problem in the Ashtekar
theory. Thus, following the gauge fixing idea leads to some
delicate problems.

How to overcome these problems will be discussed in separate
papers, see \cite{CKR}, \cite{JKR}. Instead of trying to fix the
gauge, one rather has to find a generating set of genuine
invariants. Below we show that it is quite easy, to write down a
highly redundant set of invariants, but it is very hard to reduce
it. If we want to work with genuine invariants we are
automatically forced to consider higher order monomials, built
from basic bosonic and fermionic fields. These invariants inherit,
of course, some (anti)-commutation relations, but the algebra
generated by them does not close on the linear level. This way
interesting new algebras occur. We refer to \cite{JR} for some
first remarks on their structure. It turns out that algebras of
similar types have been discussed in different areas of
mathematical physics throughout the last decade, see the list of
references in \cite{JR}.

Since the generators of ${\mathfrak O}^{\infty}_{\Lambda}$ have
been already listed before, it remains to discuss ${\mathfrak
O}^i_{\Lambda}$ in terms of generators and relations.


\subsection{Generators of ${\mathfrak O}^i_{\Lambda}$}
\label{generatorsobsalg}


Below, we define a set of generators of ${\mathfrak
O}^i_{\Lambda}$ in terms of gauge-invariant combinations of the
fields $(U,E,\psi,\psi^* )$. In the next subsections, we will
systematically reduce the number of generators to a minimal set.

\begin{Theorem}
\label{generators} The observable algebra ${\mathfrak
O}^i_{\Lambda}$ is generated by the following gauge invariant
elements (together with their conjugates):
\begin{eqnarray}
   \label{observableU}
   U_{\gamma} & := & U^A_{\gamma \, A} \, \\
   \label{observableE} E_{\gamma}(x,y) & := & U^A_{\gamma \, B} \,
   {E^B}_A(x,y) \, \\
   \label{observableJ} J^{a b}_{\gamma}(x,y) & :=
   & {\psi^{*a}}_{A}(x) \, U^A_{\gamma \, B} \, \psi^{bB}(y) \, \\
   \label{observableW} W^{a b c}_{\alpha \beta \gamma}(x,y,z) &:= &
   \tfrac{1}{6} \epsilon_{ABC} \, U^A_{\alpha \, D} \, U^B_{\beta \,
   E} \, U^C_{\gamma \, F} \, \psi^{aD}(x) \,\psi^{bE}(y)
   \,\psi^{cF}(z) \, ,
\end{eqnarray}
with $\gamma$ denoting an arbitrary closed lattice path in formula
(\ref{observableU}), a closed lattice path starting and ending at
$x$ in (\ref{observableE}) and a path from $x$ to $y$ in
(\ref{observableJ}). In formula (\ref{observableW}), $\alpha$,
$\beta$ and $\gamma$ are paths starting at some reference point
$t$ and ending at $x$, $y$ and $z$, respectively. In formula
(\ref{observableE}), both $x$ and $y$ stand also for $\infty$.
\end{Theorem}
For the proof see Appendix \ref{Casimirs}.

Note that the observables $J^{a b}_{\gamma}$ and $W^{a b
c}_{\alpha \beta \gamma}$ represent hadronic matter of mesonic and
baryonic type. They will play a basic role in future
investigations towards a construction of an effective theory of
interacting hadrons. Basically, the lattice hamiltonian can be
expressed in terms of the above invariants. In particular, the
kinetic energy $E^2$ of the gluonic field is given by second
Casimirs, its potential energy $B^2$ by Wilson loops $U_{\gamma}$
and the matter field part is given in terms of $J$'s, (which,
however, are related with $W$'s {\em via} non-linear constraints).


\subsection{The reduction idea}
\label{reduction idea}

The above generating set turns out to be highly redundant. There
is a number of non-trivial relations between generators, inherited
from the canonical (anti)- commutation relations and from the
local Gauss laws. Below, we will show how to solve the local Gauss
laws explicitly. This will be done by using a technique, based
upon the choice of a {\em lattice tree}. This way we shall prove
that ${\mathfrak O}^i_{\Lambda}$ can be decomposed (in a
tree-dependent way) into the tensor product of a gluonic and a
matter field part. This presentation of ${\mathfrak
O}^i_{\Lambda}$ can be constructed in two steps:
\begin{enumerate}
\item First we fix a lattice point $x_0$ and impose gauge
invariance with respect to the pointed gauge group at $x_0$,
\begin{equation}
  \label{pointedgaugegroup} G^0_{\Lambda} = G^{i,0}_{\Lambda}
  \times G_\Lambda^{\infty}
\end{equation}
with
\begin{equation}
  \label{pointedgaugegroup1} G^{i,0}_{\Lambda} = \prod_{x_0 \neq x
  \in \Lambda^0 } G_x \, .
\end{equation}
Moreover, we implement the Gauss laws at all points ${\Lambda}^0
\ni x \ne x_0$, i.~e.~we factorize with respect to the ideal
${\mathfrak I}^0 _{\Lambda} \cap (G^0_{\Lambda})'$, with
${\mathfrak I}^0 _{\Lambda}$ being generated by the Lie algebra
${\mathfrak g}^{i,0}_{\Lambda} \subset {\mathfrak
g}^{i}_{\Lambda}$ of $G^{i,0}_{\Lambda}$. This gives the pointed
algebra of internal observables:
\begin{equation}
  \label{observ1}
{\mathfrak O}^{i,0}_{\Lambda } := (G^0_{\Lambda})'
  / \{{\mathfrak I}^0_{ \Lambda} \cap
  (G^0_{\Lambda})'\} \, .
\end{equation}
\item Next, we impose on ${\mathfrak O}^{i,0}_{\Lambda}$ gauge
invariance with respect to the residual gauge group $G_{x_0}$, and
factorize with respect to the ideal ${\mathfrak I}_{x_0} \subset
{\mathfrak O}^{i,0}_{\Lambda}$ generated by the local Gauss law at
$x_0$.
\end{enumerate}

Whereas the first step can be performed without any obstructions,
in the second step, all the problems mentioned at the beginning of
this chapter show up.

\begin{Theorem}
The internal observable algebra can be viewed as follows:
\begin{equation}
  \label{obs-ost} {\mathfrak O}^i_{\Lambda} \cong (G_{x_0})' /
  \{{\mathfrak I}_{x_0} \cap  (G_{x_0})' \} \, ,
\end{equation}
where $(G_{x_0})'$ and ${\mathfrak I}_{x_0}$ are considered as
subalgebras of ${\mathfrak O}^{i,0}_{\Lambda}$.
\end{Theorem}

\noindent {\bf Proof:} Since $G_{x_0}$ commutes with
$G^0_{\Lambda}$, we have $(G_{\Lambda})' = (G^0_{\Lambda})' \cap
(G_{x_0})'$. Moreover, the invariant subspaces $H^0$ and $H_{x_0}$
of $G^{i,0}_{\Lambda}$ and $G_{x_0}$ are both closed subspaces of
$H_\Lambda$, whereas ${\mathfrak I}^0_{\Lambda}$ and ${\mathfrak
I}_{x_0}$ are composed of operators vanishing on $H^0$ and
$H_{x_0}$, respectively. Observe that
$$
H^0 \cap H_{x_0} = {\cal H}_\Lambda
$$
and that the ideal
$$
{\mathfrak I}_{\Lambda} =
{\mathfrak I}^0_{\Lambda} \oplus {\mathfrak I}_{x_0}
$$
is composed of those operators which vanish on this intersection.
Using completely analogous arguments as in the proof of Theorem
(\ref{structurobsalgebra}), we obtain
\begin{equation}
   \label{pointedalgebra}
   {\mathfrak O}^{i,0}_{\Lambda } =
   {\mathfrak K}(H^0) \cap (G_\Lambda^{\infty})' \ .
\end{equation}
In the second step we have to factorize the commutant $(G_{x_0})'$
of $G_{x_0}$ in ${\mathfrak O}^{i,0}_{\Lambda }$ with respect to
${\mathfrak I}_{x_0} \cap (G_{x_0})' \, .$ But by
(\ref{pointedalgebra}) we have
$$
  (G_{x_0})' = (G_{x_0})'({\mathfrak K}(H^0))
  \cap (G_\Lambda^{\infty})'
$$
where $(G_{x_0})'({\mathfrak K}(H^0))$ is the commutant taken in
${\mathfrak K}(H^0)$. Again, using similar arguments as in the
proof of Theorem (\ref{structurobsalgebra}), we get
$$
 (G_{x_0})' / \{ {\mathfrak I}_{x_0} \cap (G_{x_0})' \}
\cong {\mathfrak K}({\cal H}_\Lambda) \cap (G_\Lambda^{\infty})',
$$ which is isomorphic to ${\mathfrak O}^i_{\Lambda }$, by formula
(\ref{O-calH}). \qed


\subsection{Reduction with respect to pointed gauge transformations}
\label{pointedalgebra1}


As already mentioned, a convenient way to solve relations between
generators is to choose a {\em tree}, i.~e.~to assign a unique
path connecting any pair of lattice sites. More precisely, a tree
is a pair~$(x_0, \tree)$, where $x_0$ is a distinguished lattice
site (called {\em root}) and $\tree$ is a set of lattice links
such that for any lattice site $x$ there is exactly one path from
$x$ to $x_0$, with links belonging to~$\tree$. We denote this path
by $\beta(x)$. Consequently, for any pair $(x,y)$ of lattice
sites, there is a unique {\em along tree} path from $x$ to $y$,
equal to $\beta^{-1}(y) \circ \beta(x)$, where $\alpha \circ
\beta$ denotes the composition of the two paths (a path obtained
by first running through $\beta$ and next through $\alpha$) and
$\beta^{-1}$ denotes the path taken with the opposite orientation.
This does apply to sites at infinity also, because {\em external
links} are treated as belonging to the tree {\em a priori} .

To find an explicit set of generators of the pointed observable
algebra ${\mathfrak O}^{i,0}_{\Lambda}$, given by
(\ref{pointedalgebra}), we ``parallel transport'' all generators
of the field algebra to the lattice root using the above {\em
along tree} paths. The transported generators feel only gauge
transformations at $x_0$ and, therefore, are invariant with
respect to $G^{i,0}_\Lambda$. Hence, $ (G^0_{\Lambda})^\prime$ is
generated by:
\begin{enumerate}
\item
\begin{equation}
\label{set}
  \left\{U_{\gamma}^A{}_B, E_{\tree}^A{}_B(x,y),
  \psi_{\tree}^{aA} (x),\psi_{\tree}^{*aA} (x) \right\} \, ,
\end{equation}
where $\gamma$ is an arbitrary closed curve starting and ending at
$x_0$ and
\begin{eqnarray}
\label{pointed--U} E_{\tree}^A{}_B(x,y) & := &
U_{\beta(x)}^A{}_C \, U_{\beta(x)^{-1}}^D{}_B \,
{E}^C{}_D(x,y) \,
 , \\ \psi_{\tree}^{aA} (x) & := &
U_{\beta(x)}^A{}_B \, \psi^{aB} (x) \, ,
 \end{eqnarray}
with $\beta(x)$ denoting the unique tree path from $x$ to $x_0$.
\item
boundary fluxes:
\begin{equation}
\label{E-treeb}
   E_{\tree}^A{}_B(x,\infty) :=
   U_{\beta(x)}^A{}_C \, U_{\beta(x)^{-1}}^D{}_B
   \, {E}^C{}_D(x,\infty) \,.
\end{equation}

\end{enumerate}
The generating set (\ref{set}) is still enormously redundant. To
reduce this redundancy, in a first step, we restrict the
admissible paths to the form:
\begin{equation}
\label{special-path}
  \gamma(x,y) := \beta(x) \circ (x,y) \circ
  \beta^{-1}(y) \ .
\end{equation}
We denote
\[
U_{\tree}^A{}_{B}(x,y):= U_{\gamma(x,y)}^A{}_B \ .
\]
It is obvious that any $U_{\gamma}^A{}_B$ may be reconstructed
from those quantities. Thus,
\begin{equation}
\label{set1}
  \left\{U_{\tree}^A{}_{B}(x,y),
  E_{\tree}^A{}_B(x, y),
  \psi_{\tree}^{aA} (x), \psi_{\tree}^{*aA} (x)
  \right\}
\end{equation}
can be taken, together with boundary fluxes, as a set of
generators of $(G^0_{\Lambda})'$. The bosonic and fermionic
generators fulfill the same commutation relations as generators
$U^A{}_{B}(x,y)$ and ${E}^A{}_B(x, y)$ (see Section
\ref{FieldAlgebra}). The local Gauss law can be easily rewritten:
\begin{equation}
 \label{Gauss-1}   {\rho_{\tree}^A}_B(x) =
 \sum_{y\leftrightarrow x} E_{\tree}^A{}_B(x, y) \ ,
\end{equation}
where $\rho_{\tree}$ is given by (\ref{rho}) with $\psi^{aA} (x)$
replaced by $\psi_{\tree}^{aA} (x)$. However, a nontrivial
commutator between $E_{\tree}^A{}_B(x, y)$ and the latter occurs.

Next, observe that for any on-tree-link $(x,y) \in \tree$ we have
$$ U_{\tree}^A{}_{B}(x,y) = \delta^A{}_B \, . $$ Thus, the
relevant information is carried by those $U_{\tree}$'s, which
correspond to off-tree-links $(x,y) \notin \tree$. On the other
hand, exactly those among the fields $E_{\tree}$, which correspond
to off-tree-links, may be chosen as independent generators.
Indeed, the on-tree $E_{\tree}$'s can be calculated by solving the
local Gauss law at all the points $x \ne x_0$.  Observe that the
off-tree $E_{\tree}$'s have trivial commutators with
$\psi_{\tree}$'s. Thus, factorization of $(G^0_{\Lambda})'$ with
respect to the local Gauss laws at all points $x \neq x_0$
consists in taking only independent internal generators,
i.~e.~those among (\ref{set1}), which correspond to
off-tree-links.

Let us denote the number of lattice sites by $N$ and by $L$ the
number of links. Since the number of on-tree links is equal to
$N-1$, the number of off-tree links is equal to $$ K = L - N + 1
\, . $$ Enumerate these links putting $(x_i,y_i)=:\ell_i$, where
$i=1,\dots , K$. We have thus the following set of independent
generators of ${\mathfrak O}^{i,0}_{\Lambda}$:
\begin{equation}
\label{set2}
  \left\{U_{\tree}^A{}_{B}(\ell_i), E_{\tree}^A{}_B(\ell_i),
  \psi_{\tree}^{aA} (x), \psi_{\tree}^{*aA} (x)
  \right\} \ ,
\end{equation}
together with the boundary fluxes (\ref{E-treeb}). The latter
commute with all generators (\ref{set2}) and, after the final
reduction with respect to $G_{x_0}$, they will generate the center
of the algebra.

Generators (\ref{set2}) fulfill canonical (anti)-commutation
relations, given by formulae (\ref{CCR2}),(\ref{CCR1}) and
(\ref{CCR3}). They are all subject to gauge transformations at the
tree root $x_0$. Observe that, since bosonic and fermionic
generators commute, we have
\begin{equation}
\label{tensor0}
   {\mathfrak O}^{i,0}_{\Lambda} = {\tilde{\mathfrak O}}^{glu}_{\tree}
   \otimes {\tilde{ \mathfrak O}}^{mat}_{\tree}
   \otimes {\mathfrak O}^b_\Lambda \ ,
\end{equation}
where ${\tilde{\mathfrak O}}^{glu}_{\tree}$ is the gluonic part
generated by $\left\{U_{\tree}^A{}_{B}(\ell_i),
E_{\tree}^A{}_B(\ell_i) \right\}$ and ${\tilde{ \mathfrak
O}}^{mat}_{\tree}$ is the matter field part, generated by
$\left\{\psi_{\tree}^{aA} (x), \psi_{\tree}^{*aA} (x) \right\}$.
According to the above discussion, the gluonic algebra
${\tilde{\mathfrak O}}^{glu}_{\tree}$ is isomorphic to the
generalized CCR-algebra over the group G, spanned by $K$ pairs of
generators, and ${\tilde{ \mathfrak O}}^{mat}_{\tree} $ is
isomorphic to the CAR-algebra, generated by $12N$ pairs of
anticommuting elements. The subalgebra ${\mathfrak O}^b_\Lambda$
denotes the component generated by boundary fluxes
(\ref{E-treeb}).


\subsection{Removing the residual gauge freedom}
\label{residualgauge}


In the first part of this paper we have
mentioned that $G$ can be, basically, an arbitrary compact Lie
group. Here, we definitely consider  $G = SU(3)$ only. In what
follows, we denote the $K$-fold cartesian product of $G$ by ${\bf
G}^K = G\times \dots \times G $ and elements of ${\bf G}^K$ by
${\bf g} = (g_1, \dots, g_K) \, .$

Let us denote the off-tree variables by
$$
E_i = E_{\tree}{}(\ell_i) \, \, \, , \, \, \, U_i
= U_{\tree}(\ell_i) \, , \, \, i = 1, \dots, K \, .
$$
The residual gauge group $G_{x_0} \cong G$ acts on this set of
variables by
\begin{equation}
\label{homo} (E_i,U_i) \rightarrow (g E_i g^{-1} , g U_i g^{-1})
\,, \,
\end{equation}
with $g = g(x_0) \in G_{x_0} \, .$ In what follows, we want to fix
this residual gauge freedom. Thus, we have to consider the action
of $G$ on ${\bf G}^K$ by inner automorphisms
$$
G \times {\bf G}^K   \ni (h, (g_1, \dots , g_K) ) \mapsto
(h g_1 h^{-1}, \dots , h g_K h^{-1}) \in {\bf G}^K \,.
$$
We wish to parameterize, by choosing a gauge, the space of
equivalence classes of elements of ${\bf G}^K$ with respect to
this group action, which by abuse of language, will be called
${\rm Ad}G\, .$ Factorizing with respect to this action, we obtain
the orbit space ${\bf G}^K/{\rm Ad}G$. This is a complicated
stratified set, which will be more deeply discussed in \cite{CKR}.
Here, we restrict ourselves to the generic orbit type,
respectively the generic stratum ${\bf G}^K_{gen} \, ,$ which is
an open and dense submanifold in ${\bf G}^K \, .$ An element ${\bf
g} = (g_1, \dots, g_K) \in {\bf G}^K$ belongs to the generic
stratum, iff its stabilizer is the center ${\mathbb Z}_3$ of $G\,
.$ It is quite obvious that ${\bf g}$ belongs to the generic
stratum, iff there does not exist any common eigenvector of the
matrices $(g_1, \dots, g_K)\, .$ Moreover, one can show \cite{CKR}
that ${\bf g}$ belongs to the generic stratum, iff there exists a
pair $(g_i,g_j)$ or a triple $(g_i,g_j,g_k)$ of elements not
possessing any common eigenvector. Using arguments developed in
\cite{Fleisch} one can prove that the bundle
$$
\pi \colon  {\bf G}^K_{gen} \to  {\bf  G}^K_{gen}/{\rm Ad}G
$$
is non-trivial, for $K \geq 2 \, .$ It can be considered as a
principal fibre bundle with structure group $G/{\mathbb Z}_3 \, .$
Moreover, one can find a system of local trivializations
(respectively local sections) of this bundle, defined over a
covering of ${\bf G}^K_{gen}/{\rm Ad}G$ with open subsets, which
are all dense with respect to the natural measure (the one induced
by the Haar-measure).

Thus, let
$$
{\bf G}^K_{gen}/{\rm Ad}G \supset {\cal U}
\ni [{\bf g}] \rightarrow {\bf s}([{\bf g}]) \equiv
(s_1, \dots, s_K)([{\bf g}])
\in {\bf G}^K
$$
be one of these local sections, with ${\cal U}$ being dense in
${\bf G}^K_{gen}/{\rm Ad}G\, .$ Since $Ad G$ acts (pointwise) on
this section, we can fix the gauge by bringing ${\bf s}$ to a
special form. Since pairs of group elements being in a non-generic
position form a set of measure zero in ${\bf G}^2$, we can --
without loss of generality -- assume that $s_{K-1}$ and $s_K$ are
in generic position on ${\cal U} \, .$ That means they have no
common eigenvector. Thus, on this neighbourhood, we can fix the
gauge in two steps: First, we diagonalize $s_{K-1}$ and next we
use the stabilizer of this diagonal element to bring $s_K$ to a
special form. Since $s_{K-1}$ and $s_K$ have no common
eigenvector, this fixes the (remaining) stabilizer gauge
completely, (up to ${\mathbb Z}_3$ ). Let us denote the function
(which obviously depends only on $s_{K-1}$ and $s_K$) implementing
this gauge transformation by
\[
\pi^{-1}({\cal U}) \ni (s_1, \dots , s_K) \mapsto f(s_1, \dots ,
s_K) = f(s_{K-1} , s_K) \in G
\]
and the local section after gauge fixing by
\begin{equation}
\label{gauge-fixing}
      {\bf G}^K_{gen}/{\rm Ad}G \supset {\cal U}
  \ni [{\bf g}] \rightarrow {\bf f}([{\bf g}])
  \equiv (f_1, \dots, f_K)([{\bf g}])
  \in {\bf G}^K \, ,
\end{equation}
with
$$
f_i = f(s_{K-1} , s_K) \cdot s_i \cdot f(s_{K-1} , s_K)^{-1}
\, , \quad i =1, \dots ,K  \, .
$$
The section ${\bf f}$ can be made explicit by using a system of
local trivializations of $G$ as an $SU(2)$-principal bundle over
$S^5 \, .$ We refer to \cite{CKR} for details and to Appendix
\ref{orbits} for one example of a local section of this bundle.

Suppose that we had started with another section $\tilde {\bf s}$,
related to ${\bf s}$ by a gauge transformation given by $g \in G
\, .$ Then, the function $\tilde f(s_{K-1} , s_K)  = f(s_{K-1} ,
s_K)\cdot g^{-1}$ yields the same section $\bf f\, .$ Thus, $f$ is
equivariant with respect to gauge transformations,
\begin{equation}
\label{gh}
f(g \cdot s_{K-1} \cdot g^{-1},g \cdot s_K \cdot g^{-1})
= f(s_{K-1} , s_K) \cdot g^{-1} \, ,
\end{equation}
and in this sense, we can consider the $f_i$ as being ``gauge
invariant''. It is challenging to parameterize classes $[{\bf g}]$
of gauge equivalent configurations more intrinsically, namely in
terms of genuine invariants. In \cite{CKR} we will prove the
following

\begin{Theorem}
\label{invariants}
Any function on ${\bf G}^2$ invariant with
respect to the action by inner automorphisms
$$
G \times {\bf G}^2   \ni (h, (g_1,g_2) ) \mapsto
(h g_1 h^{-1},h g_2 h^{-1}) \in {\bf G}^2
$$
can be expressed as a function in the following invariants and
their complex conjugates:
\begin{eqnarray}
   T_1(g_1,g_2) & := & tr(g_1),\nonumber\\
   T_2(g_1,g_2) & := &  tr(g_2),\nonumber\\
   T_3(g_1,g_2) & := & tr(g_1g_2),\nonumber\\
   T_4(g_1,g_2) & := & tr(g_1{g_2}^2),\nonumber\\
   T_5(g_1,g_2) & := & tr({g_1}^2 {g_2}^2 g_1 g_2 )-
   tr({g_1}^2 g_2 g_1 {g_2}^2).\nonumber
\end{eqnarray}
Moreover, there is one algebraic relation between those invariants
such that for given values of $T_i$, $i=1\ldots 4$, there are at
most two possible values of $T_5$.
\end{Theorem}
By this Theorem, it follows that the entries of $f_i \, , \,  i =
K-1, K \, ,$ and, therefore, the group elements $f_i$ themselves
can be expressed in terms of the above set of invariants:
$$
f_i= f_i(T_1( g_{K-1},g_K) , \dots , T_5 (g_{K-1},g_K) )
\, , \quad i = K-1, K \, .
$$
Since the section $\bf f$ parameterizes the gauge orbit space, it
is clear that the remaining group elements $f_i$, $i = 1,\dots ,
K-2$, can be expressed as a function of traces
(\ref{observableU}), too.

To summarize, applying this special gauge-fixing to a gauge
configuration $(U_1, \dots, U_K) \, ,$ with generic
$(U_{K-1},U_{K})$ corresponding to the pair $(x_{K-1},y_{K-1}),
(x_{K},y_{K})$ of off-tree-links, we obtain a local
parameterization of its gauge equivalence class by
$({\mathfrak{u}}_1, \dots ,{\mathfrak{u}}_K)$ defined by
\begin{equation}
\label{u}
  {\mathfrak{u}}_i :=
  f(U_{K-1} , U_K)\cdot U_i \cdot (f(U_{K-1} , U_K))^{-1} \ .
\end{equation}
We denote the indices of these matrices by $r, s, ... \, ,$
${\mathfrak{u}}_i = \left\{ {{{\mathfrak{u}}_i}^r}_s \right\}\, ,$
and, consequently, the matrix elements of $f$ by $f = \left\{
{{f}^r}_A \right\}\, .$ Then we have
\begin{equation}
\label{urs}
   {\mathfrak{u}}_i^r{}_{s}
   = {{f}^r}_A \cdot  U_i^A{}_{B}
   \cdot {(f^{-1})^B}_s
\end{equation}
and the gauge transformation (\ref{gh}) reads:
\begin{equation}
\label{gauge-h}
  {{f}^r}_B  \rightarrow  {{f}^r}_A \cdot
  {{(g^{-1})}^A}_B \ ,
\end{equation}
(the colour indices $A,B,C \dots$ feel gauge transformations,
whereas indices $r,s, \dots$ -- assuming the same values $1,2,3$,
-- label ``gauge-invariant quantities'').

By inspecting formula (\ref{specialform}), we see that there are
two independent degrees of freedom in the matrix
${\mathfrak{u}}_{K-1}$ and 6 independent degrees of freedom in the
matrix ${\mathfrak{u}}_K$. They may be combined into 8 degrees of
freedom of a single element of $G$ in the following way:
\begin{equation}\label{uN}
  {\mathfrak{u}}_0 := {\mathfrak{u}}_{K-1} \cdot {\mathfrak{u}}_K \cdot
 \left( {\mathfrak{u}}_{K-1}\right)^{-1}
   \ .
\end{equation}
The above procedure is defined up to a discrete symmetry only.
This symmetry arises from the action of the permutation group $S_3
\subset {\rm Ad} SU(3)$ on entries of the diagonal matrix
${\mathfrak{u}}_{K-1} \in T^2 \subset SU(3)$. To fix this
$S_3$-gauge freedom means choosing for each element of  $T^2 /
S_3$ a unique representative on the torus $T^2 \, .$  Changing
this representative by an even permutation does not change the
image of the mapping $({\mathfrak{u}}_{K-1}, {\mathfrak{u}}_K )
\mapsto {\mathfrak{u}}_0$ given by (\ref{uN}), which {\em does not
cover} the entire group $SU(3)$ but only `` half of it''. Using
also odd permutations we cover a dense subset of $SU(3)$, but then
the discrete symmetry ${\mathfrak{u}}_0 \rightarrow
{\overline{\mathfrak{u}}}_0$ must be taken into account.

It can be shown that the decomposition of ${\mathfrak{u}}_0$ into
the above product of elements of special form is unique (up to the
discrete symmetry) and both ${\mathfrak{u}}_{K-1}$ and
${\mathfrak{u}}_K$ may be reconstructed from ${\mathfrak{u}}_0$.
By the above discussion, we can consider the observable
${\mathfrak{u}}_0$ as $SU(3)$-valued, provided we keep the
discrete symmetry ${\mathfrak{u}}_0 \rightarrow
{\overline{\mathfrak{u}}}_0$. To summarize, we have shown that,
locally, the full information carried by the fields $U_i$ is
encoded in $K-1$ elements ${\mathfrak{u}}_i$, $i = 0, \dots ,
K-2$, of $G$, modulo the discrete symmetry just described.

Analogously, we construct $K$ gauge invariant generators
\begin{equation}
\label{e}
   {\mathfrak{e}}_i^r{}_s
   = {{f}^r}_A \cdot  E_i^A{}_{B}
   \cdot {(f^{-1})^B}_s \ ,
\end{equation}
which have to fulfill the residual Gauss law at $x_0$. To describe
the unconstrained information carried by the fields
${\mathfrak{e}}_i$, we divide the information contained in
${\mathfrak{e}}_{K-1}$ and ${\mathfrak{e}}_{K}$ (16 gauge
invariant generators) into $8$ independent generators encoded in
the momentum ${\mathfrak{e}}_0$ canonically conjugate to
${\mathfrak{u}}_0$ and $8$ other combinations of
${\mathfrak{e}}_{K-1}$ and ${\mathfrak{e}}_{K}$, which can be
reconstructed from the global Gauss law at $x_0$. More precisely,
at each point of the section (\ref{gauge-fixing}), we decompose
the pair $({\mathfrak{e}}_{K-1},{\mathfrak{e}}_{K})$ into a pair
$({\mathfrak{e}}_{K-1}^\|,{\mathfrak{e}}_{K}^\|) $ of vectors
tangent to this section and a pair
$({\mathfrak{e}}_{K-1}^\perp,{\mathfrak{e}}_{K}^\perp)$ of vectors
orthogonal to it. Here, orthogonality is of course meant in the
sense of the natural scalar product induced by the Killing metric.
The tangent components sum up to the momentum ${\mathfrak{e}}_0$
canonically conjugate to ${\mathfrak{u}}_0 \, .$ More precisely,
${\mathfrak{e}}_0$ is the image of
$({\mathfrak{e}}_{K-1}^\|,{\mathfrak{e}}_{K}^\|) $ under the
tangent mapping of $({\mathfrak{u}}_{K-1}, {\mathfrak{u}}_K )
\mapsto {\mathfrak{u}}_0$ given by (\ref{uN}). By a simple
calculation, we get:
\begin{equation}
  \label{e--zero}
  {\mathfrak{e}}_0 = \left(\mbox{\rm Ad}
  {\mathfrak{u}}_0^{-1} - 1 \right) \circ \mbox{\rm Ad}
  {\mathfrak{u}}_{K-1}({\mathfrak{e}}_{K-1}^\|) + \mbox{\rm Ad}
  {\mathfrak{u}}_{K-1}({\mathfrak{e}}_{K}^\|) \, ,
\end{equation}
This formula is invertible and enables us to calculate uniquely
both ${\mathfrak{e}}_{K-1}^\|$ and ${\mathfrak{e}}_{K}^\|$ once we
know ${\mathfrak{e}}_0$. On the other hand, the sum
${\mathfrak{e}}_{K-1} + {\mathfrak{e}}_{K}$ is given from the
Gauss law. This enables us to calculate
$({\mathfrak{u}}_{K-1},{\mathfrak{u}}_{K},{\mathfrak{e}}_{K-1},
{\mathfrak{e}}_{K})$ once we know $({\mathfrak{u}}_0 ,
{\mathfrak{e}}_0)$.  We end up with $2(K-1)$ independent
generators $ ({\mathfrak{e}}_i , {\mathfrak{u}}_i ) \, \, , \, \,
i = 0, \dots , K-2  \, , $ of the gluonic part of the observable
algebra.

It is easy to show that these bosonic generators satisfy the
generalized canonical commutation relations over $G :$
\begin{eqnarray}
\label{e-e}
  \left[ {\mathfrak{e}}_i^r{}_s  ,{\mathfrak{e}}_j^p{}_q
  \right]
  &=&  \delta_{ij}\left({\delta^p}_s {\mathfrak{e}}_i^r{}_q  -
  {\delta^r}_q {\mathfrak{e}}_i^p{}_s \right)  \ , \\
  \label{e-u}
  \left[{\mathfrak{e}}_i^r{}_s ,{\mathfrak{u}}_j^p{}_q
  \right]
  & = &  \delta_{ij}
    \left({\delta^p}_s {\mathfrak{u}}_i^r{}_q - \tfrac{1}{3}
    {\delta^r}_s {\mathfrak{u}}_i^p{}_q \right) \, ,
    \\
    \left[{\mathfrak{u}}_i^r{}_s ,{\mathfrak{u}}_j^p{}_q
  \right]
  & = & 0
    \ .\label{u-u}
\end{eqnarray}

For the fermionic observables, we denote:
\begin{equation}
\label{ai}
  {\mathfrak{a}}^{ar}(x) :=  {{f}^r}_A
  \psi_{\tree}^{aA} (x) =  {{f}^r}_A U_{\beta(x)}^A{}_B \,
  \psi^{aB} (x)  \ .
\end{equation}
Introducing the joint index $k=(a,r,x) \, , \ k = 1, \dots ,12N ,$
we get,
\begin{equation}
\label{ak}
  {\mathfrak{a}}_k := {\mathfrak{a}}^{ar}(x) \ .
\end{equation}
Formally, these quantities fulfil the canonical anti-commutation
relations
\begin{equation}
\label{anti-comm--a} \left[{\mathfrak{a}}^{*k} ,{\mathfrak{a}}_l
\right]_+ = \delta^k{}_l \ ,
\end{equation}
but again, an additional discrete symmetry has to be taken into
account. This symmetry arises, because the section ${\bf f}\,,$
given by (\ref{gauge-fixing}), is defined only up to the
stabilizer  ${\mathbb Z}_3$ of the generic stratum. Observe that
this ambiguity does not affect the bosonic quantities
${\mathfrak{u}}$ and ${\mathfrak{e}}\,,$ because they are
``quadratic'' in $f\, .$

Let us denote the bosonic (resp. fermionic) observable algebra
part, obtained from ${\tilde{\mathfrak O}}^{glu}_{\tree}$ (resp.
${\tilde{ \mathfrak O}}^{mat}_{\tree}$) after fixing the residual
gauge, by ${\mathfrak O}^{glu}_{\tree}$ (resp. ${\mathfrak
O}^{mat}_{\tree}$). Then we have
\begin{equation}
\label{tensor}
   {\mathfrak O}_\Lambda  = {\mathfrak O}^{glu}_{\tree} \otimes
   {\mathfrak O}^{mat}_{\tree} \otimes {\mathfrak O}^b_\Lambda
   \otimes {\mathfrak O}^\infty_\Lambda  \ .
\end{equation}
The above discussion shows that locally (on a dense subset of the
generic stratum) and up to discrete symmetries, ${\mathfrak
O}^{glu}_{\tree}$ (resp. ${\mathfrak O}^{mat}_{\tree}$) coincides
with the algebra of generalized canonical commutation (resp.
anti-commutation) relations for the reduced data
$({\mathfrak{u}}_i,{\mathfrak{e}}_i)\,,$ with $ i = 0, \dots , K-2
\, ,$ (resp. $({\mathfrak{a}}_k, {\mathfrak{a}}^{*k}) \, ,$ with
$k = 1, \dots ,12N $).

As already mentioned at the beginning of this section, a
systematic study of ${\mathfrak O}_\Lambda$ as an algebra defined
in terms of generators and relations on the level of genuine
invariants will be presented in \cite{CKR} and \cite{JKR}. In
particular, we will show that the fermionic part is generated by
the following sesqui-linear and trilinear combinations of
${\mathfrak{a}}_k$ and ${\mathfrak{a}}^{*k}$:
\begin{eqnarray}
  \mathfrak{j}^k{}_l & = & {\mathfrak{a}}^*{}^k \, {\mathfrak{a}}_l
  \, , \label{observ-a1}\\
  \mathfrak{w}_{pqr} & = & {\mathfrak{a}}_p
  \, {\mathfrak{a}}_q \, {\mathfrak{a}}_r \, , \label{observ-a2} \\
  {\mathfrak{w}}^*{}^{ijk} & = & {\mathfrak{a}}^*{}^k \,
  {\mathfrak{a}}^*{}^j \, {\mathfrak{a}}^*{}^i \,. \label{observ-a3}
\end{eqnarray}
Similarly, to parameterize the bosonic part in terms of genuine
invariants, one has to take -- according to classical invariant
theory -- all trace-invariants, built from  ${\mathfrak{e}}$ and
${\mathfrak{u}}\, .$ This set is, however, highly redundant and it
is a complicated task to find, for a fixed number $K$, the full
set of relations.

\newpage

\begin{appendix}
\renewcommand{\theequation}{\Alph{section}.\arabic{equation}}

\setcounter{equation}{0}
\section{Appendix: A Local Parameterization of \\
$(SU(3) \times SU(3))_{gen}/ {\rm Ad}SU(3)$}
\label{orbits}


Consider the action of the group of inner automorphisms, here
denoted by ${\rm Ad}SU(3)$, on $SU(3) \times SU(3)$. In Subsection
\ref{residualgauge} we have used an explicit local
parameterization of the generic stratum of this action in terms of
a bundle section. Such a section can be obtained as follows. Let
$(g_1,g_2) \in SU(3) \times SU(3)$ be a pair of group elements
lying in the generic stratum. First, we diagonolize $g_1$. Since
$g_1$ is generic, the stabilizer of the ${\rm Ad}$-action is
isomorphic to $U(1) \times U(1) \, .$ Next, treating $SU(3)$ as an
$SU(2)$-principal bundle over $S^5 \, ,$ one can bring $g_2$ to a
special form using the $U(1) \times U(1)$-action. This yields a
family of local sections defined on dense subsets over the generic
stratum, corresponding to a family of local trivializations of the
$SU(2)$-principal bundle $SU(3) \rightarrow S^5 \, .$ For details
we refer to \cite{CKR}.

As a local section of the above type one can choose:
\begin{equation}
\label{specialform}
f_1 = \left[
\begin{array}{ccc}
\lambda_1 & 0 &0 \\
 0 &\lambda_2& 0 \\
 0 & 0 &\lambda_3\\
\end{array}
\right], \quad f_2 = \left[
\renewcommand{\arraystretch}{1.6}
\begin{array}{c|cc}
a & -\delta^{-1}b^{\dag} \\ \hline b & \delta \left({\mathbbm 1} -
\frac{bb^{\dag}}{1+|a|} \right)
\end{array}
\renewcommand{\arraystretch}{1}
\right] \times \left[
\begin{array}{c|rr}
1 &0  & 0 \\ \hline
 0&c&d\\
 0&-\bar{d}&\bar{c}\\
\end{array}\right] \, .
\end{equation}
Here, $\lambda_i$ are eigenvalues of $g_1$, fulfilling $$
|\lambda_1|=|\lambda_2|=|\lambda_3| = 1,\quad
\lambda_1\lambda_2\lambda_3=1, $$ The entries $$ a, \delta \in
{\mathbb C} \, , \, b=\left[
\begin{array}{c}
b_1\\b_2
\end{array} \right] \, ,\, b_1,b_2 \in \mathbb{R}_+  \, , \,
$$ of the first factor in $f_2$ fulfil $$ |a|^2+b_1^2+b_2^2=1 \, ,
\, |\delta| = 1 \, , \,  a=|a|\delta^{-2} $$ and the lower
diagonal block of the second factor is an $SU(2)$-matrix in the
standard parameterization, with $$ |c|^2+|d|^2=1. $$


\setcounter{equation}{0}
\section{Appendix: Proof of Theorem \ref{generators}}
\label{Casimirs}

It is well-known that the only gauge invariant combinations built
exclusively from the $U$`s are the Wilson-loops
(\ref{observableU}).

Any other invariant is built by contracting the colour indices of
the fields ${E^A}_B(x,y)$, ${U^A}_B(x,y)$, $\psi^{aA}(x)$ and
${\psi^{*}}_{aA}(x)$. Consider such an invariant $I$ and replace
in its definition the above fields by their gauge invariant
counterparts ${\mathfrak{e}}$, ${\mathfrak{u}}$, ${\mathfrak{a}}$
and ${\mathfrak{a}}^{*}\, .$ In particular, the missing on-tree
quantities ${\mathfrak{e}}$ and ${\mathfrak{u}}$, are defined as
combinations of the off-tree ones, using the Gauss law (for
${\mathfrak{e}}$) and the Bianchi identities (for
${\mathfrak{u}}$). Formally, the new invariant obtained this way
coincides with $I$, because the factors $f$ and $f^{-1}$ coming
from the definition of the quantities ${\mathfrak{e}}$,
${\mathfrak{u}}$, ${\mathfrak{a}}$ and ${\mathfrak{a}}^{*}$
disappear under contraction. Moreover, all the fermionic
quantities ${\mathfrak{a}}$ and ${\mathfrak{a}}^{*}$ appearing in
the invariant may be grouped to give quantities ${\mathfrak{j}}$,
${\mathfrak{w}}$ and ${\mathfrak{w}}^*$.

As already mentioned in Subsection \ref{residualgauge}, the
invariant quantities ${\mathfrak{u}}$ can be expressed as
(nonlinear) functions of traces of $U$ (invariants
(\ref{observableU})). We show that also ${\mathfrak{e}}$,
${\mathfrak{j}}$, ${\mathfrak{w}}$ and ${\mathfrak{w}}^*$ can be
expressed in terms of invariants (\ref{observableU}) --
(\ref{observableW}), listed in Theorem \ref{generators}.
Invariants  ${\mathfrak{e}}_i^r{}_s$ can be dealt with as follows:
We contract them with $8$ different ${\mathfrak{u}}$`s and use
formulae (\ref{urs}) and (\ref{e}) to obtain the following system
of linear equations for the $8$ independent components of
${\mathfrak{e}}_i^r{}_s:$
$$
{\mathfrak{e}}_i^r{}_s \, {\mathfrak{u}}^{(1)s}{}_r = E_{\gamma_1} \, ,
\, \dots \, , \, {\mathfrak{e}}_i^r{}_s \, {\mathfrak{u}}^{(8)s}{}_r
= E_{\gamma_8} \, ,
$$
with the right-hand-sides all being invariants of type
(\ref{observableE}). Analogously, we can write down systems of
linear equations of this type for the quantities ${\mathfrak{j}}$,
${\mathfrak{w}}$ and ${\mathfrak{w}}^*$. Solving these systems of
linear equations, we obtain ${\mathfrak{e}}$, ${\mathfrak{u}}$,
${\mathfrak{j}}$, ${\mathfrak{w}}$ and ${\mathfrak{w}}^*$ as
functions, linear with respect to invariants (\ref{observableE}),
(\ref{observableJ}), (\ref{observableW}) and nonlinear with
respect to invariants (\ref{observableU}) listed in Theorem
\ref{generators}.

Hence, we have formally expressed $I$ as a combination $\tilde I$
of invariants listed in Theorem \ref{generators}. In particular
all Casimir operators, built from the electric fields $E$ may be
expressed in terms of these generators.

The above formulae, expressing any gauge invariant field $I$ as a
combination $\tilde I$ of the invariants listed in Theorem
\ref{generators}, were derived by the help of the gauge fixing
section (\ref{gauge-fixing}), which is not globally defined.
Hence, equality $I =\tilde I$ holds on a dense subset of the
configuration space only. But $I$ is a differential operator (with
smooth coefficients) on the whole configuration space (the rank of
such an operator is equal to its algebraic order with respect to
variables $E$). The invariant $\tilde I$ is a differential
operator of the same rank, but {\em a priori} its coefficients are
well defined on a dense set of the configuration space only. But,
if two such operators coincide on a dense set, they coincide
everywhere.

\end{appendix}

\section*{Acknowledgments}

The authors are very much indebted to K.~Schm\"udgen and
S.~L.~Woronowicz for a lot of helpful discussions and remarks.
Moreover, they are grateful to Sz.~Charzynski, P.~D.~Jarvis and
M.~Schmidt for discussions and for reading the manuscript. This
research was partly supported by the Polish Ministry of Scientific
Research and Information Technology under grant
PBZ-MIN-008/P03/2003. One of the authors (J.~K.) is grateful to
Professor E.~Zeidler for his hospitality at the Max Planck
Institute for Mathematics in the Sciences, Leipzig, Germany. The
other author (G.~R.) is grateful to the Foundation for Polish
Science (FNP) and to the Zygmunt Zaleski Foundation for their
support.


\end{document}